\documentclass[aoas]{imsart}

\usepackage{graphicx}
\usepackage[textwidth=8em,textsize=small]{todonotes}
\usepackage{amsmath}
\usepackage{amssymb}
\usepackage{amsthm}
\usepackage{amsfonts}  
\usepackage{url}
\usepackage{hyperref}
\usepackage{cleveref}
\usepackage{nicefrac}       
\usepackage{microtype}      
\usepackage{lipsum}
\usepackage{graphicx}
\usepackage[shortlabels]{enumitem}
\usepackage{mathtools}
\usepackage{stmaryrd} 
\usepackage{nicefrac} 
\usepackage{pifont}
\usepackage{natbib}
\usepackage{enumitem} 
\usepackage{sidecap}
\usepackage{algorithm}
\usepackage{algpseudocode}

\newcommand{\EE}{\ensuremath{\mathbb{E}}}
\newcommand{\bs}[1]{\boldsymbol{#1}}
\newcommand{\pop}{\ensuremath{\mathcal{X}}}
\newcommand{\popN}{\ensuremath{\mathcal{X}_{\boldsymbol{N}}}}

\newtheorem{proposition}{Proposition}
\newtheorem{definition}{Definition}

\newtheorem*{theorem*}{Theorem}
\newtheorem*{lemma*}{Lemma}
\newtheorem{lemma}{Lemma}

\newcommand{\bN}{\ensuremath{\mathbb{N}}}

\renewcommand{\Pr}{\ensuremath{\mathbb{P}}}

\newcommand\blinded[1]{\emph{blinded for review}}

\DeclarePairedDelimiter{\ceil}{\lceil}{\rceil}

\DeclareMathOperator*{\argmax}{arg\,max}

\newcommand{\feasibleNull}{\ensuremath{\mathcal{E}_0}}
\newcommand{\boundaryNull}{\ensuremath{\mathcal{C}}}

\newcommand{\nullDist}{\ensuremath{{\allDist^0}}}

\newcommand{\altDist}{\ensuremath{\allDist^1}}
\newcommand{\allDist}{\ensuremath{\aleph_{\bs{N}}}}
\newcommand{\calI}{\ensuremath{\mathcal{I}}}
\newcommand{\calX}{\ensuremath{\mathcal{X}}}

\newcommand{\Tkt}[2]{\ensuremath{T_{{#1}}(#2)}}

\begin{document}

\title{Sequential stratified inference for the mean}
\runtitle{Sequential stratified inference}

\begin{aug}
\author[A]{\fnms{Jacob V.}~\snm{Spertus}}
\author[B]{\fnms{Mayuri}~\snm{Sridhar}}
\author[A]{\fnms{Philip}~\snm{Stark}}

\address[A]{Department of Statistics, University of California, Berkeley}
\address[B]{Massachusetts Institute of Technology, Cambridge}
\end{aug}


\begin{abstract}
    
    We develop conservative tests for the mean of a bounded population under stratified sampling and apply them to risk-limiting post-election audits.
    The tests are ``anytime valid'' under sequential sampling, allowing optional stopping in each stratum. 
    Our core method expresses a global hypothesis about the population mean as a union of intersection hypotheses describing within-stratum means.
    It tests each intersection hypothesis using a test supermartingale (TSM) combined across strata by multiplication. A $P$-value for each intersection hypothesis is the reciprocal of that TSM, and the largest $P$-value in the union is a $P$-value for the global hypothesis.
    This approach has two primary moving parts: the rule selecting which stratum to draw from next, and the form of the TSM within each stratum.
    These rules may vary over intersection hypotheses. 
    We construct intersection tests with the smallest expected sample size and propagate them to approximate an efficient global test.
   In instances that arise in auditing and other applications, its expected sample size is substantially smaller than that of previous methods.
\end{abstract}

\maketitle

\section{Introduction}
\label{sec:intro}

A ubiquitous problem in applied statistics is to make inferences about the mean $\mu$ of a population $\pop$ using a random sample.
Samples can be stratified to accommodate logistical constraints or increase efficiency. 
To draw a \textit{stratified random sample}, $\pop$ is first partitioned into disjoint strata and a uniform independent random sample (UIRS) with replacement or a simple random sample (SRS) without replacement is taken from each stratum, independently across strata.

For instance, risk-limiting post-election audits (RLAs) may stratify and sample from a trustworthy paper record of cast votes in order to provide affirmative evidence that machine-tallied election outcomes are correct---that reported winners really won \citep{stark23, stark08a}. 
RLAs test sets of null hypotheses positing that one or more reported winners actually lost.
There is strong evidence that the results are correct only if those hypotheses are rejected \citep{stark20a}. 
The corresponding tests must be \textit{finite-sample valid}---meeting their nominal significance level at any sample size---and \textit{nonparametric}---since the null hypothesis indexes an infinite-dimensional set of possible distributions. 
RLAs may be stratified for legal reasons: California law stipulates that post-election audits are conducted independently across counties (California Election Code §336.5 and §15360). 
In other cases, stratifying RLAs by voting technology \citep{ottoboniEtal18, SpertusStark2022} or by proxies of vote share reduces heterogeneity, potentially reducing the overall sample size compared to unstratified sampling.

\begin{SCfigure}
    \centering
    \includegraphics[width = 0.47 \textwidth]{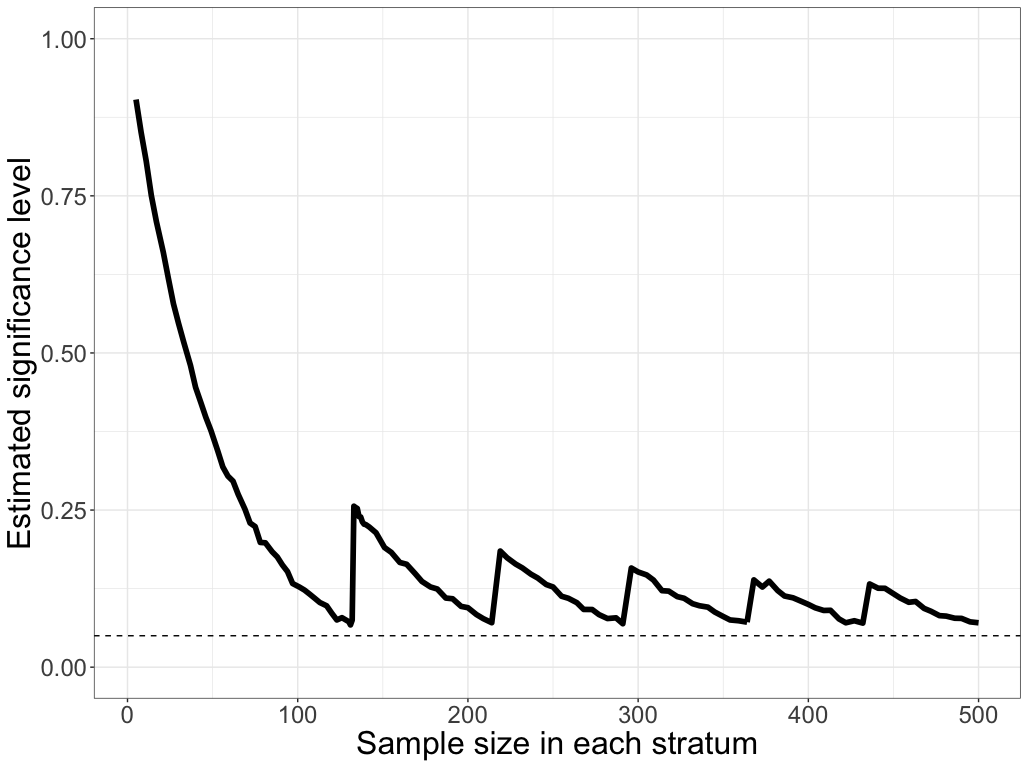}
    \caption[True significance level of the stratified Student t-test under a nonstandard distribution]{True significance level of a stratified, one-sample Student $t$-test of the (true) null hypothesis $H_0: \mu(\pop)\leq 1/2$, as a function of the sample size from each stratum.
    The strata are equal in size and have the same distribution: a mixture of a point mass at 0 (total mass 1\%) and a point mass at 0.5050505 (total mass 99\%).
    Equal-size IID samples were drawn from the two strata independently.
    The solid black line is the estimated
    significance level ($y$-axis) of the test for a range of sample sizes in each stratum ($x$-axis). 
    If the test were finite-sample valid, the solid line would never be above 5\% (dashed line), which is the nominal level of the test.
}  
    \label{fig:significance_level}
\end{SCfigure}

Stratification complicates inference. 
Texts generally suggest using a stratified version of Student's $t$-test, approximating the distribution of a weighted sum of stratumwise sample means by Student's $t$-distribution \citep{neyman34, hansenEtal53, kish65, cochran77}.
The approximation is good when $\pop$ is nearly Gaussian in every stratum, but the test can be anti-conservative when the within-stratum distributions are skewed (\Cref{fig:significance_level}).
RLA populations are often highly skewed. 
In a card-level comparison audit, individual ballot cards are compared to corresponding digital receipts of their machine interpretations. 
Most population values (corresponding to correctly interpreted ballots) lie just above the null mean with a small tail (corresponding to errors) far below.

This paper introduces methods to make conservative, non-asymptotic inferences about $\mu$ from stratified samples, without relying on parametric assumptions. 
The tests allow optional stopping and optional continuation: the analyst may check results as each sample arrives and decide whether to stop sampling and draw a conclusion or to continue gathering data.
In other words, the tests are \textit{anytime valid} over a nonparametric class of distributions.
Sequential sampling is useful for RLAs, because it is straightforward for an audit to expand as needed by retrieving more ballots from the paper trail.

When the sample is \textit{not} stratified, Wald's sequential probability ratio test (SPRT) \citep{wald45} is sequentially valid.
The SPRT was originally characterized for populations and designs that yield a parametric likelihood. 
For instance, if samples are drawn by UIRSing and the population is binary, the samples are IID Bernoulli and the population is fully parameterized by its mean.
If the null and alternative are simple, the likelihood ratio test statistic achieves the smallest expected stopping time \citep{wald45}. 
Mixing over a composite alternative provides a `test of power~1' that is guaranteed to stop when the null is false \citep{robbinsSiegmund74}.
Recent work on \textit{test supermartingales} (TSMs) and \textit{E-processes} has provided efficient sequential inference for bounded, nonparametric populations \citep{Howard21, orabona2023tight, ramdasEtal23}, interpretable in terms of `bets' against the null hypothesis and tuned so that the accumulated wealth grows quickly \citep{shafer21, grunwaldEtal24}.

A very simple approach to inference on stratified binary populations combines multiplicity-corrected confidence bounds constructed separately for each stratum \citep{wright91}. 
Wright's idea is easily generalized to bound the population mean using any set of confidence bounds that are valid within strata---the strata need not be binary. 
\citet{wendellSchmee96} improved on Wright's method for binary populations by maximizing a $P$-value over all simple hypotheses---about the stratumwise means---in a union that comprises the global hypothesis---about the pooled population mean.
Their method is intractable when there are many strata.

Our work generalizes and sharpens these methods of inference by leveraging \textit{union-of-intersection testing}: a general framework for dealing with nuisance parameters \citep{tsui89, berger94, silvapulle96} that has great utility in stratified inference. 
In brief, maximizing the $P$-value over a set of null means is equivalent to partitioning the null hypothesis into a union across all possible values of the nuisance parameters that yield the hypothesized value of the parameter of interest (a union of intersections), testing every element of the partition, and rejecting the union if and only if every simple hypothesis in the union is rejected.
This is a special case of testing a composite null, which can be handled via testing-by-betting \citep{ramdasEtal23}.
The union-of-intersections framing was suggested in \citet{ottoboniEtal18} (SUITE) for sequential inference in RLAs and by Stark in 2019 (see \url{https://github.com/pbstark/Strat})
for inference about stratified binary populations from fixed-size samples.
SUITE used Fisher's combining function to pool evidence across strata.
\citet{stark20a} generalized SUITE.
\citet{stark23} proposed (but did not evaluate) union-of-intersections tests with product combining of TSMs for inference in stratified RLAs. 
\citet{SpertusStark2022} investigated sample sizes for a range of combining functions, TSMs, and selection strategies for stratified comparison audits.

Sequential testing is also closely related to inference on multi-armed bandits (MAB).
For instance, \citet{WaudbySmith2024} provides confidence sequences for mean rewards and quantiles of contextual bandits.
\citet{Kanarios2024} formulates a \emph{cost-aware} MAB, which leads to different optimal selection strategies compared to traditional MABs. 
In stratified inference, costs vary across arms when sampling is more or less expensive depending on the stratum.
Finally, \citet{cho_peeking_2024} use betting TSMs to test composite hypotheses about weighted averages of mean rewards in MABs.
They use average combining across arms and a specialized betting rule that makes the problem computationally tractable.
To develop sharp tests for stratified sequential sampling, we use product combining across strata and optimal betting rules within strata \citep{VovkWang2021, SpertusStark2022}.

RLAs are our lead example, but financial and healthcare audits pose similar statistical problems involving highly skewed but bounded distributions \citep{fienbergEtal77, dhs20}.
For example, the United States Center for Medicare and Medicaid Services relies on random sampling \citep{cms23} to estimate and recover billions of dollars in overpayments \citep{bittinger22}.
Stratified random sampling is often employed in financial audits \citep{dhs20} and may dramatically lower the cost of the audit itself \citep{fienbergEtal77}. 
Beyond audits, our methods are applicable to sequential stratified surveys and experiments in a wide range of industrial, regulatory, and scientific applications.

The rest of this paper proceeds as follows.
\Cref{sec:notation} introduces notation and foundational concepts including sequential stratified sampling and TSMs.
We review how to test intersection hypotheses by combining marginal tests.
\Cref{sec:simple_solution} presents a basic strategy: summing confidence bounds across strata while adjusting for multiplicity.
It is unnecessarily conservative. 
\Cref{sec:uitsms} introduces union-of-intersections tests to sharpen stratified inference. 
\Cref{sec:consistency} defines notions of consistency and efficiency for sequential-stratified inference.
We describe `oracle' tests for simple alternatives, and empirical approximations to those tests that can be implemented under a composite alternative. 
We propose strategies for computation in \Cref{sec:computation}.
Some theoretically sharp methods are not computationally tractable for more than two or three strata, some work for a moderate number, and some allow for an arbitrary degree of stratification.
\Cref{sec:simulations} compares the efficiency of different strategies in simulated RLAs.
\Cref{sec:discussion} discusses our results and sketches future directions for research.
All code used to implement the methods and run simulations is available at \url{https://github.com/spertus/UI-TS}.

\section{Preliminaries and notation}
\label{sec:notation}

We use calligraphic font for sets and bags (i.e., multisets), and bold font for vectors (and sometimes for tuples).
Tuples are denoted using parentheses, e.g., $(\calX_1, \ldots, \calX_n)$; finite-dimensional vectors are denoted using square brackets, e.g., $[x_1, \ldots, x_n]$. 
If $\bs{x}$ and $\bs{y}$ are two vectors with the same dimension $K$, we write $\bs{x} \le \bs{y}$ iff $x_k - y_k \le 0$, $k=1, \ldots, K$;
and we define the dot product $\bs{x} \cdot \bs{y} := \sum_k x_k y_k$.
The vector of all zeroes is $\bs{0}$ and the vector of all 1s is $\bs{1}$, with dimension implicit from context.
The set of nonnegative integers including 0 is $\bN$.
If $\calI$ is a set or a bag, then $|\calI|$ is its cardinality.
For two scalars $a$ and $b$, $a \wedge b$ is their minimum and $a \vee b$
is their maximum. 
The operators $\Pr_{\popN}(\cdot)$ and $\EE_{\popN}(\cdot)$ denote probability and expectation induced by sampling from the population $\popN$; without the subscript, the operators are understood to be generic.
In this section, we largely define our notation without reference to a particular intersection null $\bs{\eta}$. 
When objects are conditional on the intersection null being tested, we will embellish our notation accordingly. 

\subsection{Population and parameters} 

A \textit{population} is a finite bag of real numbers $\pop := \Lbag x_i \Rbag_{i = 1}^N$. 
For simplicity, we assume\footnote{%
One-sided bounds on elements suffice to run a one-sided test or construct a one-sided confidence bound. 
Also, any lower bound on elements can be accommodated by shifting: if $x_i \ge a_i$ then $y_i := (x_i - a_i) \ge 0$.
If all elements share upper and lower bounds $a$ and $b$, they can be rescaled to $[0, 1]$ with an affine transformation $y_i := (x_i-a)/(b-a) \in [0, 1]$; the mean of the resulting population $\Lbag y_i \Rbag$ is the same affine transformation applied to the original mean.
} that each element of the population is in $[0, 1]$.
The population mean is $\mu(\pop) := N^{-1}\sum_{i = 1}^N x_i$, and we want to test a hypothesis of the form: 
\begin{equation}
    H_0: \mu(\pop)\leq \eta_0, \label{eqn:global_null}
\end{equation}
for \textit{global null mean} $\eta_0$. 
A lower $1-\alpha$
confidence bound is the largest $\eta_0$
for which $H_0$ is not rejected at level $\alpha$.
If there are upper bounds for each element of $\pop$, an upper one-sided test can be obtained 
by subtracting each element from its upper bound and then using a lower one-sided test, \emph{mutatis mutandis}.

Let 
$\bs{N} := [N_1, \ldots, N_K]$ denote the
vector of stratum sizes.
The symbol $\popN := (\pop_k)_{k=1}^K$ represents a generic stratified population---a tuple of $K$ bags with $N_k$ items in the $k$th bag $\pop_k$ 
with $N = \sum_k N_k$.
The symbol $\popN^\star := (\pop_k^\star)_{k=1}^K$ denotes the true, unknown population.
The symbol $\allDist$ represents all $K$-tuples of bags of numbers in $[0, 1]$ such that the $k$th bag, $\pop_k$, has $N_k$ items; that is, $\allDist$ denotes all stratified $[0, 1]$-valued populations with the requisite number of items in each stratum.
We use $x_{ki}$ to denote a generic element of the $k$th stratum, e.g.,
$\pop_k = \Lbag x_{ki} \Rbag_{i=1}^{N_k}$.
The vector of \textit{stratumwise means} for population $\popN$ is 
$\bs{\mu}(\popN) := [\mu(\pop_1), \ldots , \mu(\pop_K)]$, 
where $\mu(\pop_k) = \sum_{i} x_{ki} / N_k$.
The vector of \textit{stratum weights} is
$\bs{w} := [w_1, \ldots, w_k]$,
where $w_k := N_k / N$.
The mean of a stratified population $\popN$ is 
$\mu(\popN) :=  \bs{w} \cdot \bs{\mu}(\popN).$
Let $\nullDist :=
\{ \popN \in \allDist : \mu(\popN) \le \eta_0 \}$ denote the set of null populations;
the \emph{global null hypothesis} can be written $H_0: \popN^\star \in \allDist^0$.
Also let $\altDist := \{ \popN \in \allDist : \mu(\popN) > \eta_0 \}$ denote the set of alternative populations.
Together, $\nullDist$ and $\altDist$ partition $\aleph_{\bs{N}}$.

Let $\mu^\star := \mu(\popN^{\star})$ be the true global mean, and $\bs{\mu}^\star := \bs{\mu}(\popN^{\star}) = [\mu_1^\star,\ldots,\mu_K^\star]$  be the true stratumwise means.
An \emph{intersection null hypothesis} is the assertion 
$\bs{\mu}^\star \le \bs{\eta}$ for the \emph{intersection null mean} $\bs{\eta} \in [0,1]^K$. 
The global null hypothesis can be written as a union of intersection null hypotheses:
\begin{equation}
\label{eq:unionDecomposition}
    H_0: \bigcup_{\bs{\eta} \in \feasibleNull} \{\bs{\mu}^\star \leq \bs{\eta}\} \quad \iff \quad H_0: \mu^* \leq \eta_0 
\end{equation}
where
$
\label{eq:feasible_null}
    \feasibleNull := \{\bs{\zeta} : \bs{w}\cdot \bs{\zeta} \leq \eta_0,~ \bs{0} \le \bs{\zeta} \le \bs{1}\}
$
is the set of all intersection nulls for which the global null is true.

\subsection{Sampling design}


Recall that a fixed-size stratified sample produces $K$ independent batches (one for each stratum) of data in no particular order, while a sequential stratified sample is ordered within strata.
Within stratum $k$, the data is a sequence of random variables $(X_{ki})$.
When sampling without replacement, $i$ can run from 1 to $N_k$ and $(X_{ki})_{i=1}^{N_k}$ is a random permutation of the stratum values $\Lbag x_{ki} \Rbag_{k=1}^{N_k}$. 
When sampling with replacement, $i$ runs from 1 to $\infty$ and the elements of $(X_{ki})_{i=1}^\infty$ are IID. 
In either case, draws are exchangeable within each sequence and the $K$ sequences are independent of each other.

An \emph{interleaving} of samples across strata is a stochastic process $(Y_t)$ indexed by discrete time $t$;  $Y^t := (Y_i)_{i=1}^t$ is the $t$-prefix of $(Y_i)_{i \in \mathbb{N}}$. 
An interleaving is characterized by within-stratum data and a \textit{stratum selection} $S_t$:
the item in the $t$th position in the interleaving comes from stratum $S_t$.
Let $S^t := (S_i)_{i=1}^t$.
The selection $S_t$ must be \emph{predictable}---it can depend on past data $Y^{t-1}$ but not on $Y_i$ for $i \ge t$---and may involve auxiliary randomness. 
For $k \in \{ 1, \ldots, K \}$, define selection probabilities
\begin{equation}
  p_{kt} := \Pr \left ( S_t = k ~\mid~ 
 Y^{t-1}, S^{t-1} \right ),
\end{equation}
and set $\bs{p}_t :=[p_{1t}, \ldots, p_{Kt}]$, $t \in \mathbb{N}$.
If $\bs{p}_t$ has one component equal to 1 and the rest equal to zero, the stratum selection is deterministic (conditional on the past). 
We refer to $\bs{p}_t$ as the \emph{stratum selector}.
To summarize, the stratum \emph{selection} $(S_t)$ is a stochastic process taking values in $\{1, \ldots, K\}$, while the stratum \emph{selector} $(\bs{p}_t)$ 
is a vector-valued process specifying a categorical distribution for $S_t$, given the sampling history so far. 

\begin{figure}
    \centering
    \includegraphics[width=\textwidth]{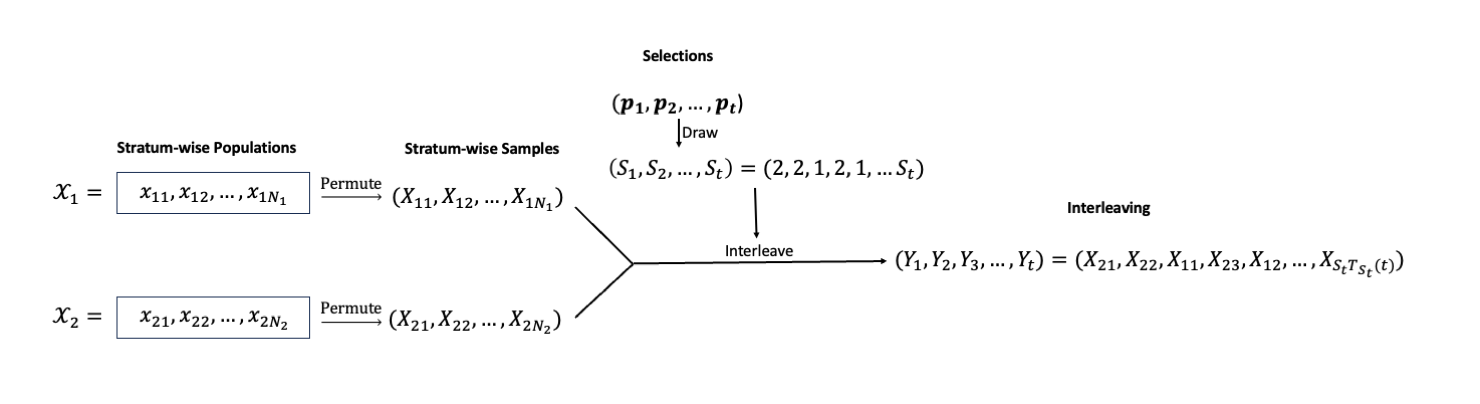}
    \caption[Sequential stratified sampling without replacement]{
    Sequential stratified sampling without replacement. 
    The population consists of $K=2$ strata, each a bag of $N_k$ numbers in $[0,1]$. 
    Within each stratum, the sampling process produces a random permutation of its corresponding bag. 
    The stratum selectors $(\bs{p}_t)$ specify the chance that draw $t$ will be from each stratum, 
    given the sampling history so far.
    The resulting selections $(S_t)$ interleave the permuted values into a single sequential sample $(Y_t)$
    for each $\bs{\eta}$.}
    \label{fig:interleaving}
\end{figure}

As of time $t$, the \textit{sampling depth} in stratum $k$ is the number of items in $Y^t$ that came from stratum $k$, denoted $\Tkt{k}{t}$. 
Thus $X_k^{\Tkt{k}{t}} := (X_{ki})_{i=1}^{\Tkt{k}{t}}$ are the data from stratum $k$ at time $t$, and the $t$th item in the interleaving,
$Y_t = X_{S_t \Tkt{S_t}{t}},$
is the $\Tkt{S_t}{t}$th item drawn from stratum $S_t$. 
We can write $Y^t = (X_{S_i \Tkt{S_i}{i}})_{i=1}^t$, and refer the reader to \Cref{fig:interleaving} for an illustration.
The selections can depend on the intersection null, in which case we augment the notation above. 
For instance, $S^t(\bs{\eta})$ is a sequence of selections for $\bs{\eta}$ and $Y^t(\bs{\eta})$ is the corresponding interleaving.
The \textit{filtration} $\mathcal{F}_t(\bs{\eta}) := \sigma(Y^t(\bs{\eta}), {S^{t+1}}(\bs{\eta}))$ is the sigma-field representing everything known up to time $t$ at null $\bs{\eta}$. 
Note that the selections are generated before the data is observed.

\subsection{Sequential hypothesis tests}

\label{subsec:hyp-test-TSM}

Valid tests follow from valid $P$-values.
\begin{definition}[Stratified sequential $P$-value]

    Let $(P_t)_{t \in \mathbb{N}}$ be a $[0, 1]$-valued stochastic process.
    Then $(P_t)$ is a sequentially valid $P$-value for the global null hypothesis $\popN \in \nullDist$ if
    for any fixed $q \in [0, 1]$,
    $$\Pr_{\popN}(\exists~ t \in \mathbb{N} : P_t \leq q ) \leq q
    \;\;\mbox{ when } \popN \in \nullDist.$$ 
\end{definition}
\noindent
If $(P_t)$ is an sequentially valid $P$-value, then rejecting the null hypothesis only when $P_t \le \alpha$ yields a sequentially valid hypothesis test with significance level $\alpha$.
The random variable 
$\tau := \inf \{t \in \bN : P_t \leq \alpha\}$ is the \textit{stopping time} of the test.

One can easily construct such $P$-values from stochastic processes that are nonnegative supermartingales under the null
\citep{shafer_test_2011}:
\begin{definition}[Test supermartingale (TSM)]
    The stochastic process $(M_t)_{t \in \mathbb{N}}$ is a 
    \emph{test supermartingale (TSM)} for $H_0$ with respect to the filtration $(\mathcal{F}_t)_{t \in \mathbb{N}}$ if, when $H_0$ is true, 
    $(M_t)_{t \in \mathbb{N}}$ is a \emph{nonnegative supermartingale starting at 1} such that $M_t \geq 0$,  $\mathbb{E}[M_t \mid \mathcal{F}_{t-1}] \leq M_{t-1}$, and $\mathbb{E}[M_0] = 1$.
\end{definition}
\noindent
TSMs yield sequentially valid $P$-values through the following result of \citet{ville39}:
\begin{proposition}[Ville, 1939]
    Let $M_t$ be a nonnegative supermartingale starting at 1.  
    For all $q \in [0,1]$, 
    $$\Pr (\exists~ t \in \mathbb{N} : M_t \geq 1/q ) \leq q.$$
\end{proposition}
\noindent 
Ville's inequality is analogous to Markov's inequality, but holds uniformly over $t \in \mathbb{N}$.
It implies that the reciprocal of the running maximum of a TSM, 
$(\max_{j \leq t} M_j)^{-1} \in [0,1]$, is a sequentially valid $P$-value for $H_0$.

\subsubsection{Constructing TSMs from random samples}

Consider a single stratum with true mean $\mu_k^\star$ and null mean $\eta_k$.
We construct a process $(M_{kt})_{t \in \bN}$ that is a TSM with respect to the samples $(X_{kt})_{t \in \bN}$ under the \emph{stratum null} $\mu_k^\star \le \eta_k$.
The \textit{conditional} stratumwise null mean $\eta_{kt}$ is the 
mean of the values remaining in $\pop_k$ at time $t$ if the null is true:
\begin{eqnarray*}
\eta_{kt} = \left \{ \begin{array}{ll}
\eta_k & \mbox{for sampling with replacement,} \\
\frac{\eta_k N_k - \sum_{i=1}^{\Tkt{k}{t-1}} X_{ki}}{N_k - \Tkt{k}{t-1}} & \mbox{for sampling without replacement.}
\end{array}
\right .
\end{eqnarray*}
Generically, a within-stratum TSM can be written 
$M_{k t_k}(\eta_k) := \prod_{i=1}^{t_k} Z_{ki}(\eta_{k})$,
where $Z_{ki}(\eta_{k})$ satisfies $\EE[Z_{ki}(\eta_{k}) \mid X_k^{i-1}] \leq 1$ when $\mu_k^\star \leq \eta_k$.
In particular, we use the \textit{betting TSM}
$$
M_{kt}(\eta_k) := \prod_{i=1}^{\Tkt{k}{t}} [1 + \lambda_{ki}(X_{ki} - \eta_{ki})].
$$
The term $\lambda_{ki} \in [0, 1/\eta_{ki}]$ is a user-defined \text{bet}, which determines how much $M_{kt}(\eta_k)$ grows or shrinks after the difference $(X_{ki} - \eta_{ki})$ is observed.
\citet{waudby-smithRamdas24,orabona2023tight} 
show that betting martingales can yield sharper inferences than other choices for $Z_{ki}(\eta_{ki})$, like exponential supermartingales \citep{Howard21}.
Within-stratum TSMs can be combined across strata to form a $P$-value for an intersection null.

\subsubsection{Intersection test supermartingales (I-TSMs)}
\label{sec:intersection_tests}

Given a particular intersection null $\bs{\eta}$, define the \textit{intersection TSM} (I-TSM):
\begin{equation*}
    M_t(\bs{\eta}) := \prod_{k=1}^K M_{k \Tkt{k}{t}}(\eta_k) = \prod_{k=1}^K \prod_{i=1}^{\Tkt{k}{t}} Z_{ki}(\eta_{k}) = \prod_{i=1}^t \tilde{Z}_i, \label{eqn:intersection_nnsm}
\end{equation*}
where $\tilde{Z}_i :=  Z_{S_i \Tkt{S_i}{i}}(\eta_{S_i})$ the $\Tkt{S_i}{i}$th term of the within-stratum TSM for stratum $S_i$. 
In other words, $(\tilde{Z}_t)$ is an interleaving defined by the selections $(S_t)$, in the same order as the data interleaving $(Y_t)$.
Because the selections are predictable, the I-TSM is indeed a TSM for the hypothesis $\bs{\mu}^* \leq \bs{\eta}$:
\begin{align*}
    \EE[M_t(\bs{\eta}) \mid \mathcal{F}_{t-1}(\bs{\eta})] 
    = M_{t-1}(\bs{\eta}) \EE\left [\tilde{Z}_t \mid X_{S_t}^{\Tkt{S_t}{t-1}} \right ]
    \leq M_{t-1}(\bs{\eta}).
\end{align*}
The second equality holds because samples from different strata are independent, and the inequality holds by construction of the within-stratum TSMs paired with the fact that $\mu_k^\star \leq \eta_k$. 
That the I-TSM is an $E$-process also follows from the fact that it can be written as a product of independent $E$-processes
(see, e.g., \citet{VovkWang2021}).
When sampling without replacement, we define $M_t(\bs{\eta}) := 1/\alpha$ if the null is certainly false in at least one stratum. 
Finally, $P_t(\bs{\eta}) := 1 / \max_{j \leq t} M_j(\bs{\eta})$ 
is a sequentially valid $P$-value for the intersection null $\bs{\eta}$. 
If an I-TSM uses betting TSMs in every stratum, it is a \textit{betting I-TSM}.

\newpage

\section{Stratified inference with TSMs}


\subsection{Simple stratified inference: combining confidence bounds}

\label{sec:simple_solution}
A lower confidence bound (LCB) for the population mean can be used to test the global null:
reject if the LCB is greater than the null. 
An LCB for the population mean can be derived by summing LCBs for each stratumwise mean using a correction for simultaneity. 
For each stratum $k$, define the sequence $(L_{kt})_{t \in \mathbb{N}}$ where
    $
    L_{kt} := \sup_{\eta \in [0,1]} \{\eta : M_{kt}(\eta) > 1/\alpha\}$
    is an LCB constructed from a TSM.
Each of these is a \textit{separately valid} LCB in the sense that 
$\sup_{k \in \{1,\ldots,K\} } \Pr(\exists~ t \in \mathbb{N} : L_{kt} > \mu_k^\star) \leq \alpha.$
Because the strata are independent, \v{S}id\'{a}k's correction\footnote{
Sharper corrections for simultaneous LCBs may be available; for instance, through the $E$-closure principle \citep{xu2025bringing}. 
In our simulation study, we compute the LCBs with no adjustment whatsoever. 
This strategy does not guarantee validity, but provides an upper bound on the efficiency of any method based on summing LCBs. 
} guarantees that the corrected LCBs 
 $\tilde{L}_{kt} := \sup_{\eta \in [0,1]} \{\eta : M_{kt}(\eta) > 1/(1-(1-\alpha)^{1/K})\}$
 are \textit{simultaneously valid}:
 $$ \Pr \left (\exists~ (t_1, \ldots, t_K) \in \mathbb{N}^K : \bigcup_{k=1}^K \{\tilde{L}_{k t_k} > \mu_k^\star\} \right ) \leq \alpha.$$
\v{S}id\'{a}k-corrected LCBs can be used for a global test as follows.


\begin{proposition}
For each $k \in \{1,\ldots,K\}$, Let $\tilde{L}_{k T_k(t)}$ be a $(1-\alpha)^{1/K}$ LCB for $\mu^\star_k$ at time $t$.
The stratum-weighted sum
    $$L_t := \sum_{k=1}^K w_k \tilde{L}_{k T_k(t)}$$
is a sequentially valid $(1-\alpha)$ LCB for $\mu^\star$, and a test that rejects $H_0$ when $L_t > \eta_0$
is a sequentially valid level-$\alpha$ test of $H_0$ with stopping time $\tau(L_t) :=
\min \{t \in \bN: L_t > \mu_0 \}$.
\label{proposition:global_LCB}
\end{proposition}
\noindent
The proof is in \Cref{app:proof_proposition_2}. 
This approach is easy to implement whenever computing the LCBs $\{L_{k T_k(t)}\}_{k=1}^K$ is straightforward and efficient;
\citet{waudby-smithRamdas24} and \citet{orabona2023tight} describe a number of possibilities.

There are two sources of slack in using \Cref{proposition:global_LCB} to test $H_0$:
(a)~The LCB method bounds each of the $K$ components of $\bs{\mu}^\star$ separately, but we only need to bound $\mu^\star = \bs{w} \cdot \bs{\mu}^\star$.
(b)~Each of the $K$ TSMs must reach $1/(1-(1-\alpha)^{1/K})$. 
The \v{S}id\'{a}k correction may be unnecessarily conservative, but even ignoring multiplicity entirely each TSM would need to be bigger than $1/\alpha$ for the (now potentially invalid) LCB method to stop. 
In contrast, to test the intersection null $\bs{\mu}^\star \le \bs{\eta}$ only requires the \textit{product} of stratumwise TSMs to reach $1/\alpha$. 
In particular, the test could reject $\bs{\eta}$ with all $K$ TSMs equal to $\alpha^{-1/K}$.
This is a much lower hurdle in each stratum, but it requires every TSM to grow (or, at least, not to shrink) for every stratum and every intersection null in the union. 
This is the goal of the test we now describe. 

\subsection{Union-of-intersections test sequence (UI-TS)}
\label{sec:uitsms}



\begin{definition}[Union-of-Intersections Test Sequence]
A stochastic process $(U_t)_{t \in \mathbb{N}}$
is a \emph{Union-of-Intersections Test Sequence} (UI-TS)
for the composite null hypothesis $H_0: \popN^\star \in \nullDist$ if for all $\popN \in \nullDist$,
$$
\Pr_{\popN} \big ( \exists~ t \in \bN: U_t \ge 1/\alpha \big ) \le \alpha.$$
Equivalently, $(U_t)$ is a UI-TS if its truncated reciprocal $P_t := 1 \wedge (1/U_t)$ is a sequentially valid $P$-value for the null $H_0$. 
\end{definition}

Recall the union-of-intersections form of the global null in \Cref{eq:unionDecomposition} and the definition of $\feasibleNull$ above. 
As shown in \Cref{sec:intersection_tests}, we can test a particular $\bs{\eta} \in \feasibleNull$ using an I-TSM.
We can reject $H_0$ if the $P$-value for every $\bs{\eta} \in \feasibleNull
$ is less than $\alpha$, i.e., if the \textit{smallest} I-TSM evaluated over $\feasibleNull$ is at least $1/\alpha$.
Therefore, 
\begin{equation*}
M_t := \min_{\bs{\eta} \in \feasibleNull} \max_{j \leq t} M_j(\bs{\eta}) = \left [\max_{\bs{\eta} \in \feasibleNull} P_t(\bs{\eta})\right ]^{-1} = \left [\max_{\bs{\eta} \in \feasibleNull}\left \{\max_{j \leq t} M_j(\bs{\eta}) \right \}^{-1} \right ]^{-1}
\end{equation*} 
is a UI-TS for $H_0$. 
When every $M_t(\bs{\eta})$ is a betting I-TSM, $M_t$ is a \textit{betting UI-TS}.

The distribution of $M_t$ is induced by the random order of the sample within each stratum, for all strata $((X_{ki}))_{k=1}^K$ and any auxiliary randomness in the bets and selections $\{(\bs{p}_t(\bs{\eta}), \bs{\lambda}_t(\bs{\eta})\}_{\bs{\eta} \in \feasibleNull}$.
Expectations over $M_t$ average uniformly over the possible sample sequences---under sampling without replacement there are up to $\prod_k N_k!$ and under sampling with replacement there are countably infinite---and over auxiliary randomness.
The smallest I-TSM at time $t$ might not be the smallest I-TSM at time $t'$.
Since selections may depend on $\bs{\eta}$, those I-TSMs 
in general involve different filtrations.
Thus it is not clear how to denote expectations of $M_t$ conditional on the past. 
For instance, if $M_t = M_t([0,1])$ and $M_{t-1} = M_{t-1}([1,0])$, the history of $M_t$ is $\mathcal{F}_t([0,1])$ and could be different than the history $\mathcal{F}_{t-1}([1,0])$ of $M_{t-1}$.
Without the running max, a UI-TS would be an $E$-process, because it would be upper bounded by a set of TSMs \citep{ramdasEtal23}. 
Even with this alteration, it cannot be written as a TSM because it does not have a single filtration.

\subsubsection{The boundary of $\feasibleNull$}
\label{sec:boundary}

Minimizing over $\feasibleNull$ is non-trivial. 
We restrict attention to within-stratum TSMs that are always monotone decreasing in $\eta_k$. 
This makes the I-TSM $M_t(\bs{\eta})$ componentwise monotone in $\bs{\eta}$. 
Consequently, the minimum of $M_t(\bs{\eta})$ occurs on the boundary of $\feasibleNull$, i.e., the set of points in the null that are closest to the alternative.

To formalize, let $\Omega_k$ be the set of all possible means $\mu_k$ in stratum $k$. 
For example, if stratum $k$ is binary, then $\Omega_k = \{0, 1/N_k, \ldots, (N_k-1)/N_k, N_k\}$.
Let $\Omega = \prod_{k=1}^K \Omega_k$ be the Cartesian product containing all possible stratumwise means $\bs{\mu}$. 
The \textit{boundary} of $\feasibleNull$ is 
$$
\mathcal{B} := \{ \bs{\eta} \in \Omega : \bs{w} \cdot \bs{\eta} \le \eta_0 \mbox{ and }   \Omega \ni 
\bs{\zeta} > \bs{\eta} \implies
\bs{w} \cdot \bs{\zeta} > \eta_0 \}.
$$
Restricting to monotone TSMs,\footnote{%
    If $\bs{\eta} \in \feasibleNull \setminus \mathcal{B}$, then there is some point $\bs{\eta}' \in \mathcal{B}$ with $\bs{\eta} < \bs{\eta}'$ for which we enforce that $M_t(\bs{\eta})$ uses the same selections and bets as $M_t(\bs{\eta}')$. 
    As result, $M_t(\bs{\eta}') < M_t(\bs{\eta})$ for all $t$ because of the monotonicity of the stratumwise TSMs in $\eta_k$.
}
the value of $\bs{\eta}$ 
that minimizes $M_t(\bs{\eta})$ is in $\mathcal{B}$.
Now define
$$
    \boundaryNull := \{\bs{\eta} : \bs{w}\cdot \bs{\eta} = \eta_0,~ \bs{0} \le \bs{\eta} \le \bs{1}\} \subset \feasibleNull.
$$
Because the I-TSMs are componentwise monotone, optimizing over the set $\boundaryNull$ 
rather than $\mathcal{B}$ gives
a conservative result, with little slack. 
The set $\boundaryNull$ is a polytope, the intersection of the unit $K$-cube with the hyperplane $\bs{w} \cdot \bs{\eta} = \eta_0$. 
This geometry influences the computational properties of UI-TSs:
in short, relaxing $\mathcal{B}$ to $\mathcal{C}$ greatly simplifies things 
(see Sections \ref{sec:computation} and \ref{app:computation} for more details).
In what follows we always assume this relaxation.

\subsubsection{Global stopping time and global sample size}
\label{sec:global_stopping_time}

We distinguish between the \emph{stopping time} of a UI-TS and its 
\emph{workload}, 
measured by the number of samples it requires to stop. 
To do so, we temporarily embellish the notation to highlight dependence on the intersection null $\bs{\mu}^\star \le \bs{\eta}$. 
For each $\bs{\eta} \in \mathcal{C}$, denote the selections by $(S_t(\bs{\eta}))_{t \in \bN}$, the selection rule by $(\bs{p}_t(\bs{\eta}))_{t \in \bN}$, and the stratumwise sample sizes by $\{T_k(t, \bs{\eta})\}_{k=1}^K$. 
The stopping time of the level $\alpha$ test induced by UI-TS $M_t$ at $\bs{\eta}$ is
$\tau(M_t; \bs{\eta}) := \min \{t \in \bN: M_t(\bs{\eta}) \ge 1/\alpha \}$.
This is the number of overall samples needed for $M_t(\bs{\eta})$ to reject $\bs{\eta}$;
and $\tau_k(M_t; \bs{\eta}) := T_k(\tau(M_t; \bs{\eta}), \bs{\eta})$ is the number of samples needed from stratum $k$. 
The \textit{global stopping time} is
the sample size needed for the ``last'' I-TSM, considered on its own, to hit or cross $1/\alpha$:
$$
    \tau(M_t) := \inf \{t \in \mathbb{N} : M_t \ge 1/\alpha\} 
    = \sup_{\bs{\eta} \in \mathcal{C}} \sum_{k=1}^K \tau_k(M_t; \bs{\eta}).
$$
On the other hand, the \textit{global sample size} $n_\tau$ is the total number of samples drawn across all strata when $H_0$ is rejected:
$$n_\tau(M_t) := \sum_{k=1}^K \sup_{\bs{\eta} \in \boundaryNull} \tau_k(M_t; \bs{\eta}).
$$
The sample size is the more relevant quality for designing a sequential stratified sample or evaluating the efficiency of a UI-TS, because it quantifies the number of physical samples required to stop. 
The stopping time bounds the sample size according to the following lemma  (the simple proof appears in \Cref{app:proof_lemma_sample_size_stopping_time}).
\begin{lemma}
    \label{lemma:sample_size_stopping_time}
    Let $M_t$ be a UI-TS for the composite null $H_0: \popN^\star \in \nullDist$.
    \begin{enumerate}
        \item For any $M_t$, potentially with selections $S_t(\bs{\eta})$ that vary across $\bs{\eta} \in \mathcal{C}$,
    $$\tau(M_t) \leq n_\tau(M_t) \leq K \times\tau(M_t).$$
        \item If $M_t$ uses the same stratum selections $S_t$ to test every $\bs{\eta} \in \mathcal{C}$,
        $\tau(M_t) = n_\tau(M_t).$
    \end{enumerate}
\end{lemma}
\noindent Below, when $M_t$ is clear from context, we will drop it from the notation.

\section{Consistency and efficiency}
\label{sec:consistency}

\subsection{Consistency}

Loosely speaking, a sequential test is consistent if it eventually rejects the global null when the global null is false. 
We can build a consistent UI-TS from I-TSMs that are \textit{intersection consistent}.



\begin{definition}[Intersection consistency]
    Consider a sequentially valid, level $\alpha$ test of the intersection null hypothesis $\bs{\mu}^\star \leq \bs{\eta}$ with $P$-value $P_t(\bs{\eta})$ at time $t$.
    If 
    $$ \mathbb{P}_{\popN}\{\exists ~t \in \mathbb{N} : P_t(\bs{\eta}) \leq \alpha\} = 1 $$
    for all $\popN$ such that $\bs{\mu}(\popN) \not \le \bs{\eta}$ and all $\alpha \in (0,1)$, then the test is \emph{intersection consistent for} $\bs{\eta}$. 
    The probability is with respect to sampling from the population and any stochastic aspects of the test, including the selection $S_t(\bs{\eta})$.
\end{definition}
\noindent
An I-TSM $M_t(\bs{\eta})$ is consistent if it grows without bound whenever $\bs{\mu}(\popN) \not \le \bs{\eta}$, producing an almost-surely shrinking $P$-value and an intersection consistent test of $\bs{\eta}$ (i.e., a `test of power 1'; see \citet[\S 3.3]{ramdas2025hypothesis}).
When sampling without replacement, consistency can be trivially satisfied by setting the I-TSM to $1/\alpha$ when the intersection null is certainly false in one stratum, even if it is certainly true in another.
When sampling with replacement, constructing a consistent I-TSM requires protecting against two failure modes.
First, in at least one stratum in which $\mu_k^\star > \eta_k$, the stratumwise TSM $M_{kt}(\eta_k)$ must grow.
Second, the stratumwise TSM for any stratum in which $\mu_k^{\star} \le \eta_k$ must not shrink towards zero without bound.
That can be accomplished by curtailing sampling in strata where there is evidence that $\mu_k^{\star} \le \eta_k$; by reducing the bets in those strata to zero; or by ensuring that at some stage, an overwhelming fraction of the samples come from some stratum in which the null is false.\footnote{%
Another strategy is to change the combining function when constructing $P_t(\bs{\eta})$. 
Averaging \citep{cho_peeking_2024} or Fisher's combining function \citep{SpertusStark2022} are consistent even when some $P$-values approach zero.  
}

\begin{definition}[Global consistency]
    Consider a sequentially valid, level $\alpha$ test of the global null $\mu^\star \le \eta_0$, and let $P_t$ denote the $P$-value for that test at time $t$.
    The test is \emph{globally consistent} iff
    $$\mathbb{P}_{\popN}\{ \exists ~t \in \mathbb{N} :P_t \leq \alpha\} = 1 $$
    whenever $\mu(\popN) > \eta_0$. 
    The probability is with respect to sampling from the population and all stochastic aspects of the test. \label{def:global_consistency}
\end{definition}
\noindent 
We will call a UI-TS \emph{consistent} if it produces a globally consistent test.  
A UI-TS must grow over the entire set of selections $\{S_t(\bs{\eta})\}_{\bs{\eta} \in \boundaryNull}$ to be consistent, and this set is included in the aforementioned stochastic aspects of the test.
In theory, we can use any set of consistent I-TSMs $\{M_t(\bs{\eta})\}_{\bs{\eta} \in \boundaryNull}$ to construct a consistent UI-TS. 
In practice, $\boundaryNull$ may be uncountably infinite, so computing the UI-TS $P$-value $P_t$ may be intractable unless the set $\{M_t(\bs{\eta})\}_{\bs{\eta} \in \boundaryNull}$ has some structure.
Perhaps surprisingly, the simplest ways to structure $\{M_t(\bs{\eta})\}_{\bs{\eta} \in \boundaryNull}$---fix the bets and selections over nulls, time, and strata---can easily yield inconsistent UI-TSs (see \Cref{app:inconsistent_uits}). 
While consistency is necessary for a useful test, it is not sufficient: a consistent UI-TS may nevertheless require an impractically large sample size.

\subsection{Efficiency}
\label{sec:efficiency}

Practically useful tests should keep $n_\tau$ relatively small when the null is false.
There is substantial subtlety lurking here because: 
(a)~the alternative usually does not specify a single distribution (a \textit{simple} alternative) but a set of distributions (a \textit{composite} alternative);
(b)~$n_\tau$ is a random variable and it is not given which features of its distribution (median, expected value, etc) should be minimized; 
(c)~for a UI-TS, $\tau$ and $n_\tau$ have complicated distributions that depend on the underlying I-TSMs.

Recall that \Cref{lemma:sample_size_stopping_time} relates the sample size and stopping time as $\tau \leq n_\tau \leq K \tau$ for general UI-TSs.
It is difficult to find the UI-TS that minimizes $\EE[n_\tau]$ over all possible bets and selections, even for a simple alternative. 
We build towards this goal by defining and relating a range of optimality properties, then constructing a UI-TS from optimal I-TSMs.
\begin{definition}[Optimality properties]
\label{def:optimality}
    Consider a betting I-TSM or UI-TS $M_t^*$ for an intersection or global null $H_0$ and fix a simple alternative $\popN \in \altDist$. 
    Let $M_t'$ be any betting I-TSM or UI-TS for $H_0$.
    \begin{enumerate}
    \item $M_t^*$ is \emph{Kelly optimal} if $\mathbb{E}_{\popN}[\log M_t^*] \geq \mathbb{E}_{\popN}[\log M_t']$ for all $t \in \mathbb{N}$. 
    \item $M_t^*$ is \emph{stopping-time optimal (STO)} if $\mathbb{E}_{\popN}[\tau(M_t^*)] \leq \mathbb{E}_{\popN}[\tau(M_t')] +1$.
    \item $M_t^*$ is \emph{efficient} if $\mathbb{E}_{\popN}[n_\tau(M_t^*)] \leq \mathbb{E}_{\popN}[n_\tau(M_t')] + 1$.
    \end{enumerate}
    We add one additional definition that pertains only to UI-TSs:
    \begin{enumerate}
    \setcounter{enumi}{3}
    \item $M^*_t$ is \emph{marginally STO} if, for every $\boldsymbol\eta \in \mathcal{C}$,
$\mathbb{E}_{\mathcal{X}_{\boldsymbol{N}}}[\tau(M^*_t; \boldsymbol\eta)]
\le \mathbb{E}_{\mathcal{X}_{\boldsymbol{N}}}[\tau(M'_t; \boldsymbol\eta)] + 1$.
    \end{enumerate}
\end{definition}
\noindent 
The slack of 1 sample in our definitions of stopping-time optimality and efficiency arises from possible overshoot of the threshold $1/\alpha$ \citep{breiman61}, which is essentially ignorable for our purposes. 
Stopping-time optimality and efficiency are equivalent for I-TSMs but not necessarily for UI-TSs, because the interleavings may vary across $\mathcal{C}$.
Below, we construct I-TSMs that satisfy all of these optimality criteria in the IID case.
We then characterize 
(a)~marginally STO UI-TSs for 
known simple alternatives;
(b)~approximately marginally STO UI-TSs for unknown simple alternatives; 
and (c)~computable approximations to (b). 
Marginally STO UI-TSs are constructed as the minimum of a collection of STO I-TSMs. 
They have no theoretical efficiency guarantees because in general we cannot relate the I-TSM stopping times to the global sample size, especially under a composite alternative. 
Nevertheless, they exhibit substantially improved empirical performance over other methods.
We also consider constraining the UI-TS to use the same selections for all $\bs{\eta} \in \feasibleNull$, so $n_\tau = \tau$ by \Cref{lemma:sample_size_stopping_time}.

\Cref{def:optimality} can be generalized in at least two ways.
First, one could use a functional of the distribution of $n_\tau$ other than its expected value, such
as a percentile of $n_\tau$, to define efficiency. 
We examine percentiles in \Cref{sec:bernoulli_simulations}.
Second, while \Cref{def:optimality} considers a simple alternative $\popN \in \altDist$, we could quantify optimality over a composite alternative $\altDist$. 
There are two standard approaches, detailed in \citet{grunwaldEtal24}: 
(1)~Summarize performance by a weighted average over the composite alternative, which is analogous to the GRO criterion;
(2)~summarize performance by the minimum efficiency over the composite alternative, which is analogous to the REGROW criterion.

We can construct a marginally STO betting UI-TS (there may be more than one) from a collection of ``greedy'' Kelly betting I-TSMs, each of which is defined as follows.

\begin{definition}[Greedy Kelly betting I-TSM]
\label{def:unconditional_ko_itsm}
    A betting I-TSM $M^*_t(\bs{\eta})$ 
    for the intersection null $\bs{\mu}^\star \le \bs{\eta}$ is Greedy Kelly for the simple alternative $\popN \in \altDist$ if its next-step expected log-growth is maximal among all betting I-TSMs for $\bs{\eta}$. 
    That is, for the filtration $\mathcal{F}_t(\bs{\eta}) := (Y^{t}(\bs{\eta}), S^{t}(\bs{\eta}))$ and all times $t \in \mathbb{N}$,
    $$\EE_{\popN}[\Delta M^*_t(\bs{\eta}) | \mathcal{F}_{t-1}(\bs{\eta})] = 
    \sup_{(\bs{\lambda}, \bs{p})} \EE_{\popN}[\Delta M_t(\bs{\eta}) | \mathcal{F}_{t-1}(\bs{\eta})],$$
    where 
    $$\Delta M_t(\bs{\eta}) := \log(M_t(\bs{\eta})/M_{t-1}(\bs{\eta})) = \log \sum_{k=1}^K 1\{S_t(\bs{\eta}) = k\} Z_{k T_k(t)}(\eta_{k})$$ 
    is the logarithmic change in the I-TSM at time $t$. 
    The supremum is over all choices of bets and selection rules $(\bs{\lambda}_t(\bs{\eta}), \bs{p}_t(\bs{\eta}))$ 
    for which $(M_t(\bs{\eta}))_{t \in \bN}$ is an I-TSM for $\bs{\eta}$. 
    The expectation is with respect to the random sampling from $\popN$ and the random stratum selections $S^t(\bs{\eta})$.
\end{definition}
\noindent
The following proposition is a classic result of \citet{kelly56} and \citet{breiman61}, paired with the fact that a Greedy Kelly I-TSM always selects a stratum with maximal growth (see \Cref{lem:kelly-optimal_I-TSM}).
\begin{proposition}[Greedy Kelly I-TSMs are efficient under IID sampling]
    \label{prop:greedy_kelly_equals_kelly_equals_sto}
    Suppose sampling is with replacement. Let $M_t^*(\bs{\eta})$ be a greedy Kelly I-TSM for the simple alternative $\popN \in \altDist$ and let $M_t'(\bs{\eta})$ be any betting I-TSM. 
    Then $M_t^*(\bs{\eta})$ satisfies 
    $$\mathbb{E}_{\popN}[\log M_t^*(\bs{\eta})] \geq \mathbb{E}_{\popN}[\log M_t'(\bs{\eta})] ~~ \mbox{for all}~~  t \in \mathbb{N},$$ and
    $\mathbb{E}_{\popN}[\tau(M_t^*(\bs{\eta}))] \geq \mathbb{E}_{\popN}[\tau(M_t'(\bs{\eta}))] - 1$. 
    That is, a greedy Kelly I-TSM is Kelly optimal, STO, and efficient when the stratum-wise samples are IID.
\end{proposition}
\noindent
When sampling without replacement, a greedy Kelly I-TSM may not be Kelly optimal because the log-growths change as strata are depleted. 
\Cref{app:greedy_counterexample} provides an example.
In fact, there is no time-uniform Kelly optimal solution in the sense of \Cref{def:optimality} because the selections that maximize the expected log wealth at time $t$ need not nest within the selections that maximize it at time $t'$.
Selections that yield an STO or fixed-time-Kelly-optimal I-TSM correspond to the optimal policy of a finite-horizon Markov bandit \citep{gittins79}, and can be found by stochastic dynamic programming (See \Cref{app:stochastic_dp}).
We emphasize that this subtlety only arises due to the optimization of selections: greedy bets are Kelly optimal and STO under any fixed sample sequence, including unstratified sampling or stratified sampling with a pre-determined interleaving (e.g. round robin).


We now identify the bets and selection rules that parameterize a Greedy Kelly I-TSM.
The optimal bets $\bs{\lambda}_t^*(\bs{\eta})$ do not depend on the selection rule $\bs{p}_t(\bs{\eta})$. 
\begin{lemma}[Constructing a Greedy Kelly I-TSM] 
\label{lem:kelly-optimal_I-TSM}
Fix an intersection null $\bs{\eta} \in \mathcal{C}$ and a simple alternative $\popN \in \altDist$. 
\begin{enumerate}
    \item \textbf{Bets:} Define the bets 
    $\bs{\lambda}_t^*(\bs{\eta}) := [\lambda_{1t}^*(\bs{\eta}), \ldots , \lambda_{Kt}^*(\bs{\eta})]$ 
    where each
    $$\lambda_{kt}^*(\bs{\eta}) := \argmax_{\lambda \in [0,1/\eta_{kt}]} \EE_{\popN} \left \{ \log (1 + \lambda (X_{kt} - \eta_{kt})) \mid \mathcal{F}_{t-1} \right \}.$$
    
    \item \textbf{Selection rules:} 
    Let $\mathcal{A}_t := \argmax_{j \in \{1, \ldots, K\}} \EE_{\popN}[\log Z_{jt}(\eta_j) | \mathcal{F}_{t-1}]$ be the set of strata with maximal expected next-step log-growth given bets $\bs{\lambda}_t(\bs{\eta})$.
    Define the selection rule 
    $\bs{p}^*_t(\bs{\eta}) := [p_{1t}^*(\bs{\eta}), \ldots, p_{Kt}^*(\bs{\eta})],$
    where
    $p_{kt}^*(\bs{\eta}) := 1\{k \in \mathcal{A}_t\}/|\mathcal{A}_t|$.
\end{enumerate}
    The I-TSM $M_t^*(\bs{\eta})$ that uses $\bs{\lambda}_t^*(\bs{\eta})$ and $\bs{p}_t^*(\bs{\eta})$
    is Greedy Kelly.
    By \Cref{prop:greedy_kelly_equals_kelly_equals_sto} it is Kelly-optimal and STO under sampling with replacement.
\end{lemma}
\noindent 
\Cref{lem:kelly-optimal_I-TSM} is proved in \Cref{app:kelly-optimal-TSM}.
For sampling with replacement (IID), the Kelly-optimal rules do not depend on $t$ and we may denote them $\bs{\lambda}^*(\bs{\eta})$ and $\bs{p}^*(\bs{\eta})$.
For sampling without replacement, the greedy Kelly bets are length-$N_k$ sequences in each stratum that depend on the sample order; there are up to $N_k!$ such sequences.

\begin{lemma}[Marginally STO UI-TS]
    \label{lem:global_kelly_optimal_uits}
    Suppose sampling is with replacement, fix a simple alternative $\popN \in \altDist$, and consider testing $H_0: \mu^\star \leq \eta$.
    For each $\bs{\eta} \in \mathcal{C}$, let $M_t^*(\bs{\eta})$ be a greedy Kelly betting I-TSM 
    and define the `greedy Kelly' UI-TS
    $$
    M_t^* := \min_{\bs{\eta} \in \mathcal{C}} \max_{j \leq t} M_j^*(\bs{\eta}).
    $$
    Given any betting UI-TS $M_t'$, $M_t^*$ satisfies:
    \begin{enumerate}
\item (Marginally STO) For every $\boldsymbol\eta \in \mathcal{C}$,
$\mathbb{E}_{\popN}[\tau(M^*_t;\boldsymbol\eta)] \le
\mathbb{E}_{\popN}[\tau(M'_t;\boldsymbol\eta)] + 1$.
\item (Lower bound) $\mathbb{E}_{\popN}[\tau(M'_t)] \ \ge\
\sup_{\boldsymbol\eta \in \mathcal{C}}
\mathbb{E}_{\popN}[\tau(M^*_t;\boldsymbol\eta)] - 1$.
\end{enumerate}
\end{lemma}
\noindent
\emph{Proof} Part 1 follows because a greedy Kelly I-TSM is STO at every $\bs{\eta} \in \mathcal{C}$ under sampling without replacement. 
Part 2 adds Jensen's inequality: $\mathbb{E}_{\popN}[\tau(M'_t)] = \mathbb{E}[\sup_{\bs{\eta}} \tau(M'_t; \bs{\eta})] \geq \sup_{\bs{\eta}}  \mathbb{E}[\tau(M'_t; \bs{\eta})] \geq \sup_{\bs{\eta}}  \mathbb{E}[\tau(M^*_t; \bs{\eta})]-1$. $\Box$ 

Trivially, a marginally STO UI-TS is consistent because the underlying I-TSMs are consistent.
However, a marginally STO UI-TS need not be STO because of the gap in Jensen's inequality, which depends on the full joint distribution of the stopping times $\{\tau(M_t;\bs{\eta})\}_{\bs{\eta} \in \mathcal{C}}$ and particularly its dispersion.
Generally, $\mathcal{C}$ can be partitioned into subsets according to which stratum is pulled by greedy Kelly selections at each $\bs{\eta}$ so that the stopping times are comonotone within subsets and independent across subsets. 
The independence contributes a small amount of slack that does not enter when \emph{all} the stopping times are comonotone.
See \Cref{app:jensen_gap} for an illustration with two intersection nulls. 
A second, larger performance gap arises from \Cref{lemma:sample_size_stopping_time}, which shows that a STO UI-TS may be inefficient up to a factor of $K$.

These gaps are both reduced by coupling the selections.
Indeed, when the selections are fixed to the same global sequence $S_t$ for every $\bs{\eta} \in \mathcal{C}$, we have $\tau(M_t) = n_\tau(M_t)$ and the stopping times $\{\tau(M_t;\bs{\eta})\}_{\bs{\eta} \in \mathcal{C}}$ are maximally correlated. 
Such a UI-TS can be inconsistent when $S^t$ is poorly chosen,\footnote{%
Suppose we are testing $H_0 :\mu^\star \leq 1/2$ by sampling with replacement and we set $S_t = 1$ for all $t$. 
Then the conditionally Kelly-optimal I-TSM for the null $[1,0]$ always bets 0 on a draw from stratum 1 and never samples from stratum 2: it does not grow.
} but consistency can be ensured by simple rules like greedy Kelly betting with round robin sampling.\footnote{%
The growth in strata where the null is false is positive. The growth in other strata is exactly zero.}
A more ambitious goal is to construct an optimal global selection rule, generating the most efficient UI-TS over all such sequences.
We do not know how to solve this optimization problem, 
but in \Cref{sec:simulations} we explore a greedy selection strategy that chooses a single selection sequence by targeting the last smallest I-TSM.



\subsubsection{Approximate optimality under the intersection-composite alternative}
\label{sec:approximate_optimality}

Consider a single intersection null $\bs{\eta} \in \mathcal{C}$ and suppose the intersection-composite alternative $\bs{\mu^\star} \not \leq \bs{\eta}$ is true. 
Within stratum $k$, the alternative $\pop_k$ is not fully specified, so we cannot directly apply Lemma~\ref{lem:kelly-optimal_I-TSM}. 
Instead, we use a \textit{predictable} approximation of a greedy Kelly strategy, that plugs in past data.
A direct strategy is to plug in the lagged empirical distribution to the expectations in \Cref{lem:kelly-optimal_I-TSM} and optimize them directly, but this generally requires expensive numerical optimizations at every $t$, $k$, and $\bs{\eta}$.
A faster approximation is provided by the AGRAPA bets of
\citep{waudby-smithRamdas24}, defined within stratum $k$ as
$
\lambda_{kt}^{\mbox{\tiny aG}}(\eta_k) := 0 \vee \frac{\hat{\mu}_{k(t-1)} - \eta_k}{\hat{\sigma}_{k(t-1)}^2 + (\hat{\mu}_{k(t-1)} - \eta_k)^2} \wedge \frac{c}{\eta_k},
$
where $\hat{\mu}_{k(t-1)}$ and $\hat{\sigma}_{k(t-1)}^2$ are the lagged empirical mean and variance, and $c \leq 1$ is a user-specified truncation parameter.

\subsubsection{Other betting strategies}
\label{sec:other_betting_strategies}

We propose two additional betting strategies that, when paired with fixed stratum selections $S_t := S_t(\bs{\eta})$, are more tractable computationally than AGRAPA and other approximately Kelly-optimal bets.
We present the strategies now and discuss their computational properties in the next section.

The first strategy sets
$\bs{\lambda}_t(\bs{\eta}) := \bs{\lambda}_t.$
Such bets may vary over time as a predictable function of past data, but must be identical for all I-TSMs $\{M_t(\bs{\eta})\}_{\bs{\eta} \in \mathcal{C}}$.
The predictable plug-in strategy $\lambda_{kt}^{\mbox{\tiny P}} := 1 \wedge \sqrt{2 \log(2/\alpha) / (\hat{\sigma}_{k(t-1)}^2 t \log t)}$, recommended by \citet{waudby-smithRamdas24},
is an instance of a fixed bet when truncated to $[0,1]$ rather than $[0,1/\eta_k]$.

The second strategy is 
$\lambda_{kt}^{\mbox{\tiny I}}(\bs{\eta}) := c_{kt}/\eta_k,$
where $c_{kt}$ is a predictable tuning parameter, and the ``I'' superscript stands for inverse.
An inverse bet is valid as long as $0 \leq c_{kt} \leq 1$
and $\eta_k > 0$. 
To set $c_{kt}$, note that for any given null $\eta_k$, it is sensible to bet more when the true mean $\mu_k$ is larger and the standard deviation $\sigma_k$ is smaller.
This suggests setting 
$c_{kt} := l_k \vee (\hat{\mu}_{k(t-1)} - \hat{\sigma}_{k(t-1)}) \wedge u_k,$ 
where $\hat{\mu}_{k(t-1)}$ and $\hat{\sigma}_{k(t-1)}$ are predictable estimates of the true mean and standard deviation. 
We set them to the lagged sample mean and standard deviation, though regularized or Bayesian estimates are also possible (as in, e.g., \citet{stark23}). 
The limits $l_k, u_k \in [0 ,1)$ are user-chosen truncation parameters ensuring the bets are valid.
As defaults, we recommend setting $l_k = 0.1$ and $u_k = 0.9$ so that some amount is always wagered but the I-TSMs cannot go broke. 
$\lambda_{kt}^{\mbox{\tiny I}}(\bs{\eta})$ produces a set of I-TSMs that are convex over $\bs{\eta} \in \mathcal{C}$, making the UI-TS more computationally tractable.

\section{Computation}
\label{sec:computation}

In principle, the results of \Cref{sec:approximate_optimality} allow one to construct an approximately optimal I-TSM for every $\bs{\eta} \in \mathcal{C}$;
Lemma~\ref{lem:global_kelly_optimal_uits} then constructs a marginally STO UI-TS under the alternative $\mu^\star > \eta_0$. 
In practice, this can be infeasible to implement. 
The tractability of the minimization depends on the constituent I-TSMs and how they vary over $\bs{\eta} \in \mathcal{C}$.
We classify I-TSMs accordingly.
The term \textit{$\bs{\eta}$-aware} refers to betting and selection strategies that depend on $\bs{\eta}$; \textit{$\bs{\eta}$-oblivious} refers to strategies that do not. 
An I-TSM or UI-TS is called $\bs{\eta}$-oblivious only if all its selectors and bets are $\bs{\eta}$-oblivious. 
We now describe a few strategies for computation under different population sizes and numbers of strata.

\paragraph*{Small $K$, small $N$, discrete support} 

Recall from \Cref{sec:boundary}
that $\mathcal{B}$ is the (possibly finite) boundary  of $\feasibleNull$.
If the support of $\nullDist$ is small and known (e.g., if $\popN^\star$ is known to be binary), if there are few strata (e.g., $K \leq 3$), and if the strata are small (e.g., $\max_k N_k \leq 1000$),
it is feasible to enumerate $\mathcal{B}$.
That makes it possible to use any $\bs{\eta}$-aware strategy 
$\{(\bs{\lambda}_t(\bs{\eta}), \bs{p}_t(\bs{\eta}))\}_{\bs{\eta} \in \mathcal{B}}$ and compute the UI-TS by brute-force.
See \Cref{app:computation}.

\paragraph*{Small $K$}
When the strata are large or the support is unknown, it is not possible to minimize over $\mathcal{B}$ by brute force.
In \Cref{app:computation} we show that $\bs{\eta}$-oblivious I-TSMs $M_t(\bs{\eta})$ are log-concave over $\bs{\eta} \in \mathcal{C}$. 
Thus, when bets and selections are $\bs{\eta}$-oblivious over a convex subset of $\mathcal{C}$, the I-TSMs are log-concave over that subset,
and the minimum over that subset must be attained at
one of its extreme points.
This suggests a strategy for finding the global minimum: partition $\mathcal{C}$ into $G$ convex subsets, within each of which the I-TSMs are $\bs{\eta}$-oblivious;
find the minimum over each subset by exploiting log-concavity; then find the minimum of those minima.

For instance, $\feasibleNull$ can be
partitioned into convex polygons and the I-TSMs need only be evaluated at the intersection of the vertices of the polygons with $\mathcal{C}$. 
A different selection rule and bet can be used for each polygon. 
We examine the case $K=2$, partitioning $\feasibleNull$ into $G$ `bands' defined by $G+1$ points along the line $\mathcal{C}$.
We give pseudo-code for this strategy in \Cref{alg:uits}, which appears at the end of our appendix.
The size and number of bands affects the statistical efficiency of the UI-TS and the computational burden. See \Cref{sec:point_mass_simulations}.

\paragraph*{Moderate $K$}
Setting $G = 1$ above would yield an $\bs{\eta}$-oblivious UI-TS. 
The resulting I-TSM $M_t(\bs{\eta})$ is log-concave in $\bs{\eta}$ over all of $\mathcal{C}$.
Its minimum is attained at a vertex of the polytope $\mathcal{C}$, so we can compute $M_t$ by evaluating $M_t(\bs{\eta})$ at the set of vertices of $\mathcal{C}$. 
The number of vertices is a combinatorial function of $K$.
Enumerating the vertices is tractable for any $\bs{N}$ and any population support if the number of strata is not too big (e.g. $K \leq 16$). 

\paragraph*{Arbitrary $K$}
If we use $\bs{\eta}$-oblivious selections $S_t(\bs{\eta}) := S_t$, the $\bs{\eta}$-aware inverse betting strategy $\bs{\lambda}_{t}^{\mbox{\tiny I}}(\bs{\eta})$ leads to an I-TSM $M_t(\bs{\eta})$ that is smooth and log-convex in $\bs{\eta}$. 
The minimum generally occurs on the interior of $\mathcal{C}$ and can be found numerically:
the objective and constraints are convex,
so projected gradient descent can find the UI-TS for arbitrary $\bs{N}$ and $K$.


\section{Application to risk-limiting audits}
\label{sec:simulations}

Risk-limiting audits (RLAs) leverage a trustworthy
record of the votes to guarantee that if any reported winner did not win, the chance that the 
reported outcome will be certified is at most the risk limit.
RLAs can be formulated as tests of hypotheses about the means of a collection of bounded lists of nonnegative numbers:
the reported winners really won if the mean of every list is greater than $1/2$ \citep{stark20a}.
Those hypotheses can be tested directly by sampling from the lists (\emph{ballot-polling audit}, BPA; technically ballot cards, not whole ballots, are sampled)
or indirectly by comparing elements of the list or sums of groups of elements to reference values derived from voting system reports (\emph{comparison audits});
see \citet{stark23}.
The most efficient methods compare list elements to a reference value for each ballot card (\emph{card-level comparison audits}, CCAs).

In this section, we compare the efficiency of UI-TSs and LCBs in simulations of stratified BPA and CCA RLAs. 
The simulations are idealized for simplicity of exposition:
for BPAs, we assume that every ballot card has a valid vote for one of two candidates,
while for CCAs, we assume that the reference values are error-free.
(For the general case, see \citet{stark20a}; non-votes or votes for other candidates will increase BPA sample sizes while errors in the reference values will generally increase CCA sample sizes.) 
The populations represent two extremes of bounded distributions: for any given mean, the point-mass distribution has minimum variance and the Bernoulli distribution has maximum variance. 
We expect performance on other distributions (e.g. beta, truncated Gaussian, three-point, etc) that arise in applied problems to be bracketed by the sample sizes estimated here. 

\subsection{Point masses: error-free card-level comparison audits}
\label{sec:point_mass_simulations}

Using the stratified CCA parametrization in \citet[\S 3.2]{SpertusStark2022}, we constructed populations with stratum sizes $\bs{N} = [200, 200]$ and identical values $x_{ik} := \mu_k^\star = 1/2$ for $i \in \{1, \ldots, 200\}$ and $k \in \{1, 2\}$, representing ballots with error-free reference values. 
The within stratum nulls were defined as $\eta_k := (1 - \theta_k - \bar{A}_k^c)/2$ where $\bar{A}_k^c$ was the \textit{stratumwise reported assorter mean}---the share of votes for the winner---and $\bs{\theta} \in \{\bs{\theta} : \bs{w} \cdot \bs{\theta} = 1/2, \bs{0} \leq \bs{\theta} \leq \bs{1} \}$. 
Letting $\bs{\bar{A}}^c := [\bar{A}_1^c, \bar{A}_2^c]$, the \textit{global reported assorter mean} $\bar{A}^c := \bs{w} \cdot \bs{\bar{A}}^c$ ranged between 0.51 and 0.75, corresponding to electoral margins between 2\% and 50\%. 
The gap between stratumwise assorter means was either $\bar{A}_2^c - \bar{A}_1^c = 0$ or $\bar{A}_2^c - \bar{A}_1^c = 0.5$.

For each population, we found the 
(deterministic) sample sizes for LCBs and UI-TSs.
The LCBs were constructed with no correction of the stratum-wise LCBs for multiplicity (e.g., \v{S}id\'{a}k's correction) in order to lower bound the sample sizes required by any valid LCB strategy.
The UI-TSs were constructed using the banding strategy discussed in \Cref{sec:computation},
partitioning $\mathcal{C}$ into $G \in \{1, 3, 10, 100, 500\}$ equal-length segments and evaluating the I-TSMs at the endpoints. 
Bets were AGRAPA, predictable mixture, or inverse. 
The selection rule was $\bs{\eta}$-oblivious, nonadaptive round-robin,
an $\bs{\eta}$-aware and adaptive ``predictable Kelly'' strategy using a UCB-style algorithm on every $\bs{\eta} \in \mathcal{C}$,
or an $\bs{\eta}$-oblivious ``greedy'' strategy using the UCB-style algorithm iteratively on the minimizing $\bs{\eta}$ at time $(t-1)$ to produce a single interleaving for all I-TSMs.

\begin{figure}
    \centering
    \includegraphics[width = \textwidth]{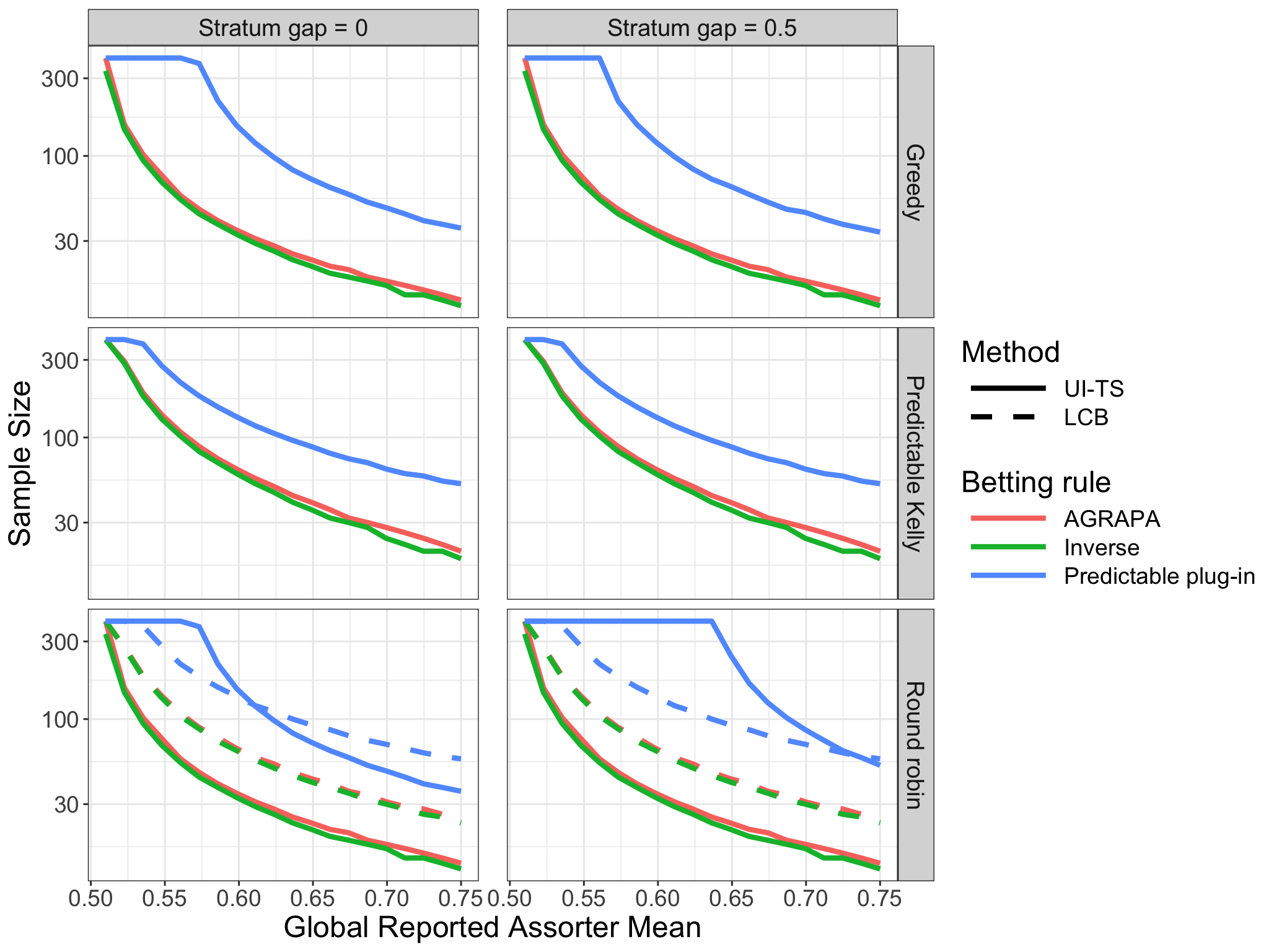}
    \caption[Sample sizes for sequential stratified tests for point-mass distributions]{Sample sizes (y-axis, log scale) for various sequential stratified tests (line colors and types) of 2-stratum, error-free, card-level comparison audit populations with varying global reported assorter means (x-axis) and spread between the within-stratum reported assorter means (columns). 
    For example, a global reported assorter mean of 0.6 and stratum gap of 0.5 means $\boldsymbol{\bar{A}}^c = [0.35, 0.85]$.
    UI-TSs were computed using the banding strategy, with $G = 100$. 
    All methods assumed sampling was with replacement, but sample sizes were capped at 400: a sample size shown as 400 is really 400 or greater. 
    LCB = lower confidence bound; UI-TS = union-of-intersections test sequence.}
    \label{fig:point_mass_results}
\end{figure}

\Cref{tab:g_sample_sizes} gives the sample sizes (averaged across populations and selection rules) for UI-TSs with AGRAPA bets for various $G$.
\Cref{fig:point_mass_results} gives sample sizes
for $G = 100$.
UI-TSs are generally more efficient than LCBs.
The difference is substantial (the y-axis is on the log scale) except when neither the bets nor selections are $\bs{\eta}$-aware (e.g., predictable plug-in bets with round robin selection). 
Despite having approximately optimal \textit{stopping times}, the predictable Kelly selection rule led to unnecessarily large \textit{sample sizes} because it used different selections for each intersection null. 
Greedy selection and round robin performed similarly, except when the stratum gap was large and the bets were fixed.

\begin{SCtable}[][ht]
\centering
\begin{tabular}{rrr}
  \hline
  $G$ & Sample size ($n_\tau$) & Run time (s) \\ 
  \hline
 1 & 400.0 & 0.3 \\ 
 3 & 239.9 & 0.7 \\ 
 10 & 92.4 & 2.0 \\ 
 100 & 57.4 & 19.8 \\ 
 500 & 54.8 & 99.0 \\ 
   \hline
\end{tabular}
\caption[Sample sizes of sequential stratified tests for different numbers of bands $G$]{Sample sizes and run times at various $G$, the number of bands for computing a product UI-TS using AGRAPA bets. 
Results are averaged across all the point-mass populations. 
The statistical efficiency of the UI-TS increases as $G$ increases, but with diminishing returns to more computation.}
\label{tab:g_sample_sizes}
\end{SCtable}

\subsection{Bernoulli distributions: ballot-polling audits}
\label{sec:bernoulli_simulations}

We created populations of size $\bs{N} = [2100, 2100]$ from Bernoulli distributions with stratumwise success probabilities $\bs{\mu^\star} = [\mu^\star - 0.25, \mu^\star + 0.25]$ for $\mu^\star \in [0.51,0.74]$. 
We recorded the sample sizes for the methods as described in \Cref{sec:point_mass_simulations}, alongside results for two additional betting strategies. 
The ``shrink-trunc Bernoulli'' strategy took the mean estimate $\hat{\mu}_{kt}$ described in \citet[\S 2.5.2]{stark23},
and transformed it into a bet as described in Section~2.3 of that paper. 
In more detail, the \emph{a priori} mean was set at $(\eta_k + 1)/2$, the anchoring factor was $d := 20$, and the estimate was truncated above $\eta_k + 1/(2\sqrt{d + T_k(t) - 1})$.
The estimate $\hat{\mu}_{kt}$ was then transformed to the bet 
$\lambda_{kt}^{\mbox{\tiny ST-Bern}}(\eta_k) = [\hat{\mu}_{k(t-1)}/\eta_k - 1] / (1 - \eta_k)$.
This is similar to the Bernoulli SPRT \citep{wald45}, but with a predictable estimate of the alternative rather than a fixed alternative.
We also implemented the Kelly-optimal bet (equivalent to Wald's SPRT for the actual alternative), which yields a lower bound on the expected stopping time of a UI-TS for any selection rule (but is not usually implementable because it requires knowing the alternative).
We estimate the expected sample size by the empirical mean sample size in 1000~simulations. 
We also estimated the power $\mathbb{P}(\tau \leq 4000)$ as the empirical rate of stopping before $t = N = 4000$.

\begin{figure}
    \centering
    \includegraphics[width = 0.49\textwidth]{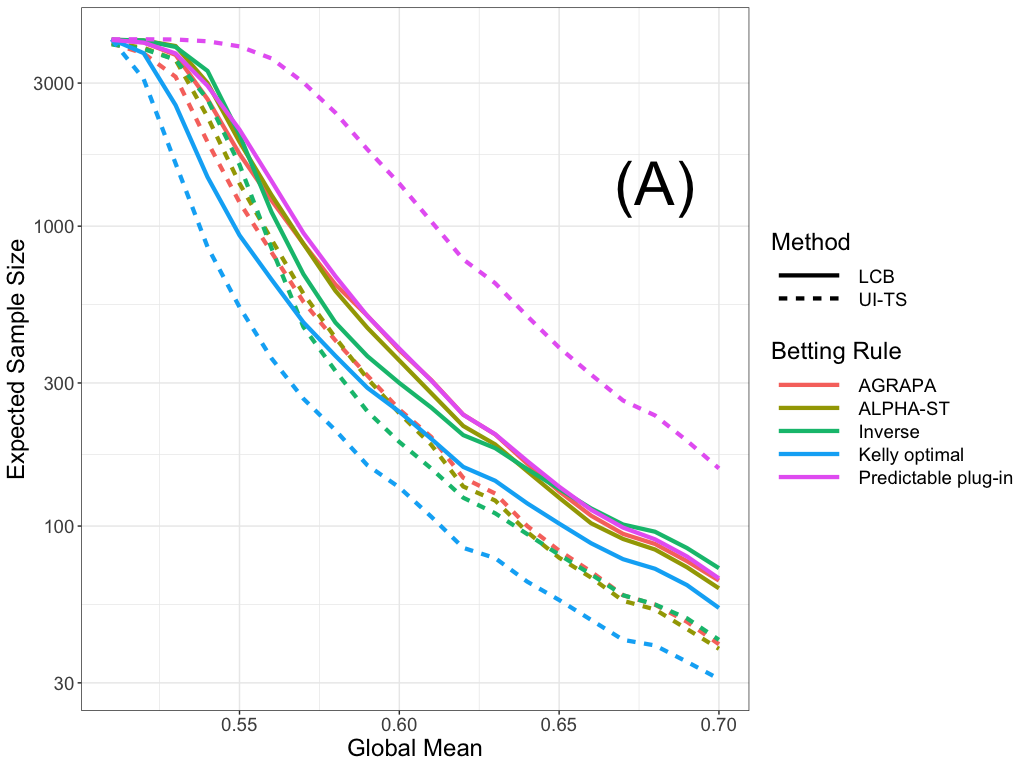}
    \includegraphics[width = 0.49\textwidth]{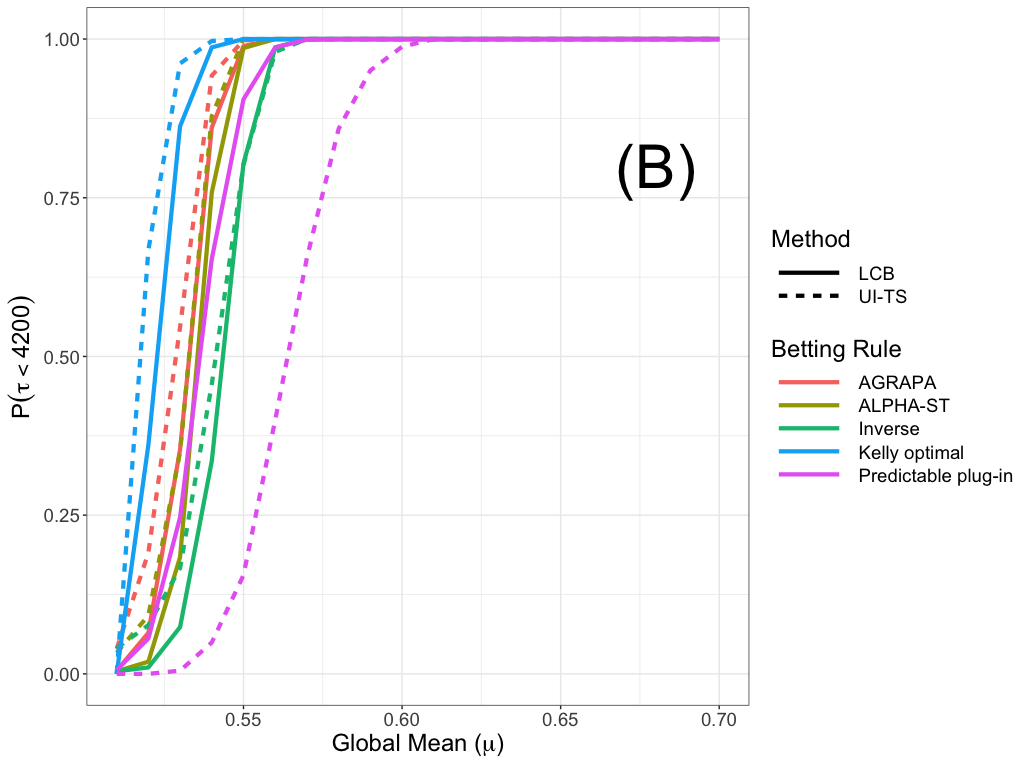}\\
    \caption[Expected sample sizes for sequential stratified tests under Bernoulli alternatives]{Expected global sample sizes $\EE[n_\tau]$ (Panel (A) y-axis; log$_{10}$ scale) and probability of stopping (Panel (B) y-axis) for the global null $H_0: \mu^\star < 0.5$ for Bernoulli populations with various true means (x-axis) and gaps between strata (columns). The LCB and UI-TS methods (linetypes) were used with various settings for the bets (line colors). The selection strategy is round robin. Expected sample sizes are taken as the empirical average sample size to stop at level $\alpha = 0.05$ over 1000 simulations. All methods assumed sampling was with replacement, but sample sizes were capped at 4000, so the expected sample size estimates may be biased downwards, in particular, when the probability of stopping is less than 1. 
    LCB = lower confidence bound; UI-TS = union-intersection test sequence.}
    \label{fig:bernoulli_sample_sizes}
\end{figure}

Results for round-robin selection appear in \Cref{fig:bernoulli_sample_sizes}.
The UI-TSs were nearly always sharper than LCBs with corresponding bets, except when predictable plug-in bets were used. 
The Kelly-optimal bets yielded the lowest expected sample size, followed by the shrink-trunc Bernoulli and AGRAPA bets. 
Inverse bets performed relatively poorly for alternatives near the global null, but improved when the true mean was large. 

\section{Discussion and conclusions}
\label{sec:discussion}

The need for sequential inference about the mean of a stratified population arises in numerous applied problems. 
We constructed tests that leverage test supermartingales (TSMs) to improve over extant methods.
The tests represent the global null as a union of intersection hypotheses each of which specifies the mean in every stratum.
The tests construct an intersection test supermartingale (I-TSM) for every intersection hypothesis, and take the smallest as a measure of evidence against the global null: a union-of-intersections test sequence (UI-TS).
UI-TSs can require substantially fewer samples than the simple approach of combining lower confidence bounds (LCBs), but their computational cost is generally higher.

We constructed marginally stopping time optimal (STO) UI-TSs from Kelly-optimal I-TSMs. 
Because STO targets stopping time (the sample size for the hardest-to-reject intersection null) rather than the overall sample size required to reject every intersection null, a (marginally) STO UI-TS is not necessarily efficient: there is typically a gap up to a factor of $K$.
We navigated this issue by evaluating selection strategies that are fixed over the union: round-robin or a ``greedy'' UCB-style selection that predictably maximizes the expected growth of the previously smallest I-TSM.
These were sharper than the marginally STO UI-TS for pointmass distributions, and that advantage will grow with $K$. 
We thus recommend using a single selection sequence for all $\bs{\eta} \in \mathcal{C}$, with approximately greedy Kelly bets. 
Open-source Python software that implements our methods and generated the figures and tables in this paper is available at
\url{https://github.com/spertus/UI-TS}.

\subsection{When stratification helps}
\label{sec:kelly_optimal_uitsm_pointmass}


Stratified sampling and inference with a UI-TS can be sharper than unstratified sampling and inference with a TSM. 
Consider a UI-TS and a TSM for a simple population: point-masses within strata (stratified sampling) or a two-point distribution (unstratified sampling).

For the UI-TS, the bets
$\lambda_{kt}^*(\bs{\eta}) = 1\{\mu^\star_k > \eta_k\} / \eta_k$
are Kelly-optimal at each $\bs{\eta}$ since the within-stratum are point-masses: 
if $x_{ik} = \mu^\star_k > \eta_k$, winning is certain, so we stake the maximum permissible amount $1/\eta_k$;
if $x_{ik} = \mu_k^\star \le \eta_k$, we cannot win, so we bet 0.
Following \Cref{lem:kelly-optimal_I-TSM}, the optimal selection rule always draws from the stratum with the largest expected log-growth. 
Because there is no randomness in the sampling, a STO I-TSM has the smallest stopping time almost surely (not merely in expectation) so a marginally STO UI-TS is STO.
The STO I-TSM at $\bs{\eta}$ is
$
M_t^*(\bs{\eta}) 
:= \max_k \left ( 
\nicefrac{\mu^\star_k}{\eta_k} \right )^t
$
and the STO UI-TS is
$ 
M_t^* = \min_{\bs{\eta} \in \boundaryNull} \max_k 
\left ( 
\mu^\star_k / \eta_k
\right )^t.
$
Specializing to $K=2$, $w_1 = w_2$, and $\eta_0 = 1/2$,
write $\eta := \eta_1$ and $(1-\eta) = \eta_2$ so that
$
M_t^* = \min_{\eta \in [0,1]} \left [(\nicefrac{\mu^\star_1}{\eta}) \vee (\nicefrac{\mu^\star_2}{1-\eta})  \right]^t
$, which can be found numerically for any $(\mu^\star_1, \mu^\star_2)$. 
The optimal stopping time is $\tau^* = \ceil{\log\alpha / (\log \eta^* - \log \mu_1^\star)}.$ 
Per \Cref{lemma:sample_size_stopping_time}, the sample size is $n_\tau(M_t^*) \leq 2\tau^*$, and in fact equality is attained: the selection rule for $[0,1] \in \mathcal{C}$ always chooses stratum 1, while the rule for $[1,0] \in \mathcal{C}$ always chooses stratum 2. 

Now take the previous example but with unstratified sampling:
each draw is equally likely to yield $\mu_1^\star$ or $\mu_2^\star$.
The expected log-growth of a betting TSM $B_t$ with bet $\lambda$ is
$
  \mathbb{E}_{\popN}[\Delta B_t] = \frac{1}{2} \log \left [1 + \lambda (\mu^\star_1 - 1/2) \right ] + \frac{1}{2} \log \left [1 + \lambda (\mu^\star_2 - 1/2) \right ]$.
When $\mu^\star_1$ and $\mu^\star_2$ are both bigger than 1/2, we are guaranteed to make money, so $\lambda^* = 1/\eta_0 = 2$.
When $\mu^\star_1$ or $\mu^\star_2$ is less than 1/2 but $\mu^\star > 1/2$, we can differentiate and solve to find $\lambda^* = 2 \wedge [(1 - \mu^\star_1 - \mu^\star_2)/(2 (\mu^\star_1 - 1/2) (\mu^\star_2 - 1/2))]$. 

\begin{figure}
    \centering
    \includegraphics[width = 0.45\textwidth]{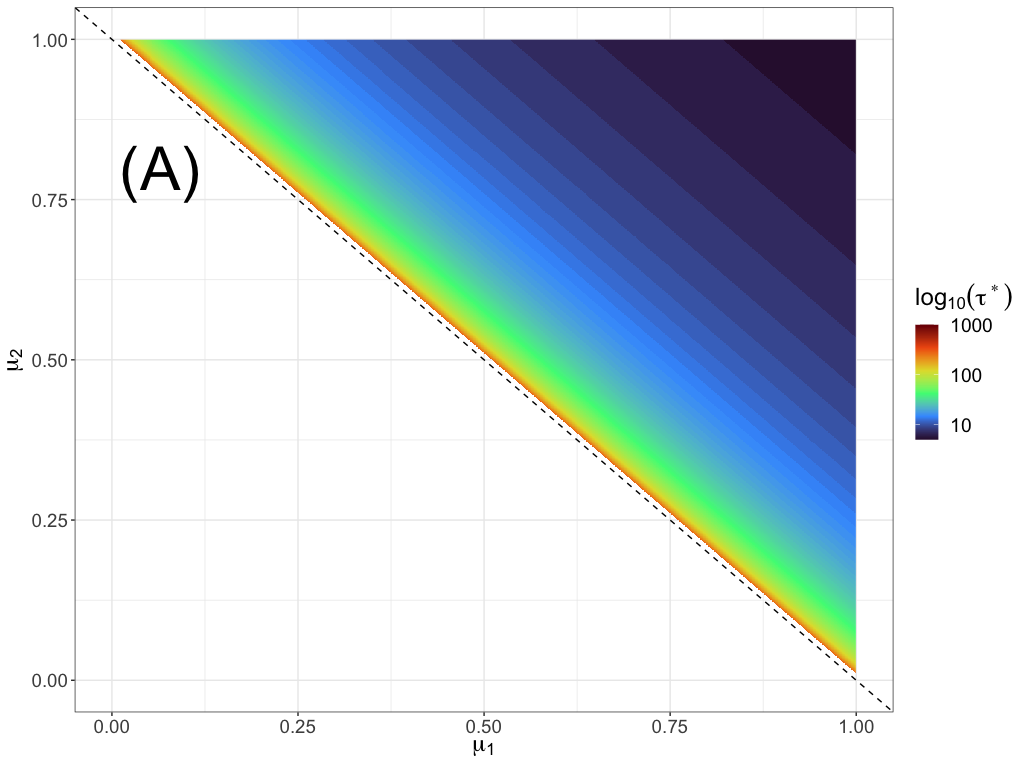}
    \includegraphics[width = 0.45\textwidth]{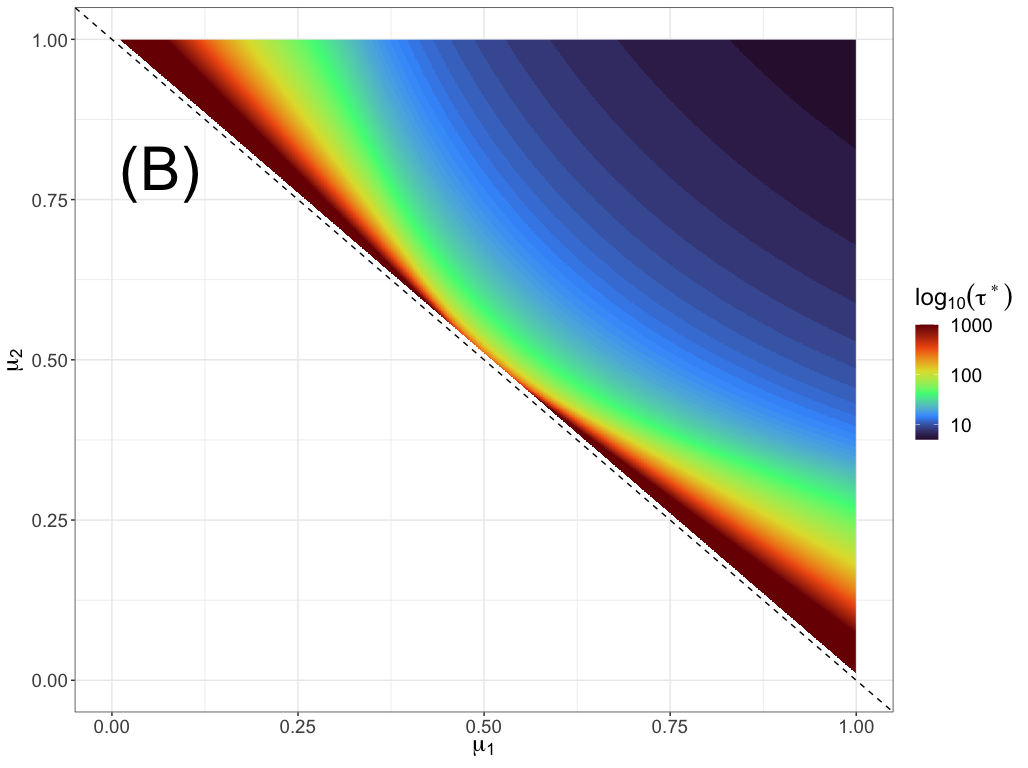}\\
    \includegraphics[width = 0.45\textwidth]{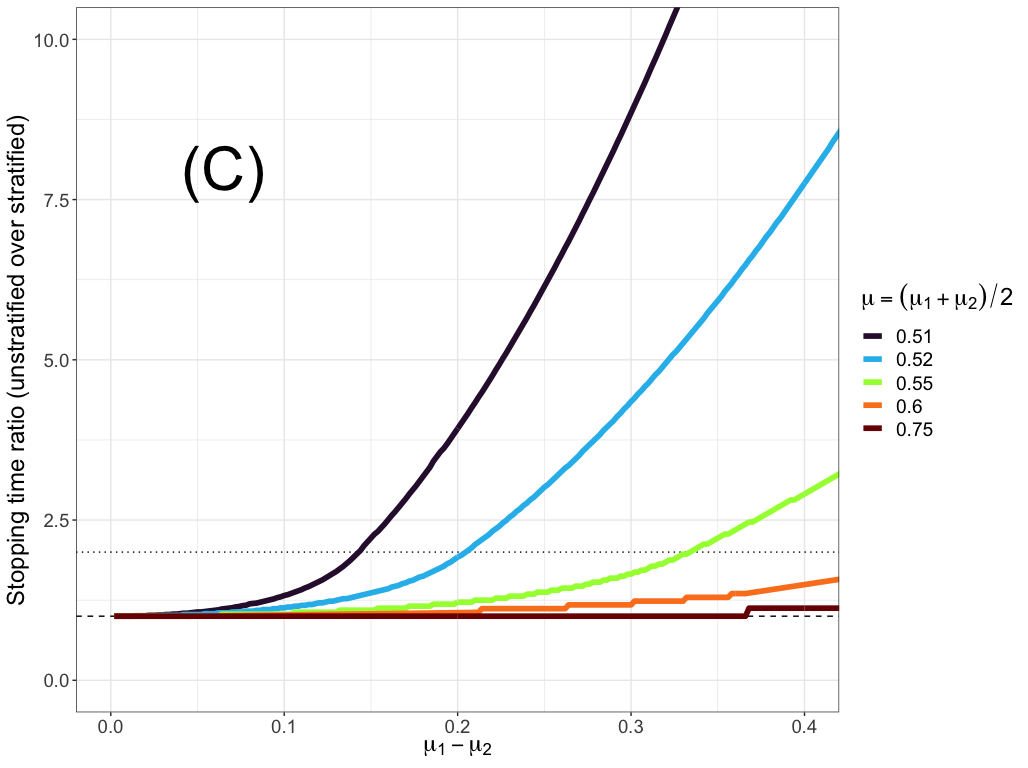}
    \caption{Stopping times of a STO stratified UI-TS and Kelly-optimal unstratified betting TSM for simple distributions. 
    Panel~(A) presents the stopping times $\tau^*$ (color, $\log_{10}$ scale) of a level $\alpha = 0.05$ test based on a STO UI-TS under sampling with replacement from a range of stratified point mass populations with means $\mu_1$ ($x$-axis) and $\mu_2$ ($y$-axis) and equal stratum weights $\bs{w} = [1/2,1/2]$. 
    Panel~(B) plots the expected stopping time (color, $\log_{10}$ scale) of a Kelly-optimal TSM under \textit{unstratified} sampling with replacement from the two-point population $\{\mu_1, \mu_2\}$.
    The color scale is capped so that dark red stopping times should be read as ``1000 or greater.''
    Panel~(C) displays the ratio of the stopping times (unstratified over stratified) for slices of $(\mu_1, \mu_2)$ defined by their difference $(\mu_1 - \mu_2$; x-axis) and average ($\mu(\pop)= \frac{1}{2}(\mu_1 + \mu_2)$; colors). 
    Stratification substantially reduces stopping times when the stratumwise means are separated, especially when $\mu(\pop)\approx 1/2$.
    Where the curves exceed 2 (dotted line) stratification reduces the global sample size as well.}
    \label{fig:kosts}
\end{figure}

The expected stopping times of both tests are given by Wald's equation and plotted in \Cref{fig:kosts}.
Stratified sampling with a STO UI-TS always beats unstratified sampling with a Kelly-optimal betting TSM. 
To translate to sample size, the UI-TS stopping times should be multiplied by 2, per \Cref{lemma:sample_size_stopping_time}. 
Stratification decreases the sample size when the margin $(\mu^\star - 1/2)$ is small and the gap between the strata $(\mu^\star_1 - \mu^\star_2)$ is large. 
This mirrors the well-known result that stratification lowers the MSE of mean estimation and the width of normal-theory asymptotic confidence intervals when within-stratum variance is low and between-stratum variance is high (e.g., \citet[\S 5.6]{cochran77}).

\subsection{Conclusions}

Finding selections that minimize the expected sample size rather than the expected stopping time is an open challenge.
We conjecture that, when the selections are constrained to be fixed across $\bs{\eta} \in \mathcal{C}$, this problem is approximately solved by greedy selection strategies like the one we examined. 

Another direction for research is to implement and evaluate methods that leverage universal portfolios \citep{cover91, orabona2023tight}. 
Relatedly, the method of mixtures could be used to construct a GRO $E$-process by assuming a prior over the alternative \citep{grunwaldEtal24}.
The concentration of the prior around the true alternative affects the efficiency but not the validity of the test. 
Applying these tools to I-TSMs could produce efficient UI-TSs. 

The UI-TSs we consider combine (independent) stratumwise $E$-values by multiplication,
which is optimal when every $E$-value is larger than~1 \citep{VovkWang2021}.
In line with that theory, we found that product combining tends to dominate other combining methods in preliminary simulation studies.
However, other ways of combining might be sharper in some cases.
For instance, Fisher's combining function negates the need for clever betting \citep{SpertusStark2022};
the average-combined, PEAK method of \citet{cho_peeking_2024} produces a consistent test;
the adaptively reweighted intersection test \citet{ekEtal23} can improve over flat averaging by updating weights online.
None of these methods leverage the efficiency of product combining. 
In simulations (\Cref{app:peak-comparison}), average combining resulted in significantly larger sample sizes: PEAK averaged about triple the sample size of our UI-TS under point-mass distributions and about 40\% higher than UI-TS for Bernoullis. 
Average combining is more robust when some $E$-processes are shrinking, but that risk is voided by Kelly betting.

It is computationally expensive to calculate a product-combined UI-TS in general. 
Partitioning $\mathcal{C}$ into convex subsets allows tuning parameters to vary while keeping the minimization tractable.
This strategy could be made adaptive, refining the partition near nulls that are challenging to reject based on the data observed so far. 
Another open problem is to maximize statistical efficiency under tractability constraints, for example by finding an efficient UI-TS while enforcing convexity over $\bs{\eta} \in \mathcal{C}$.
Elements of that class include UI-TSs that use inverse-$\bs{\eta}$ bets (\Cref{sec:other_betting_strategies}) to test every intersection null, but sharper strategies may exist.

\subsection*{Acknowledgements}
JS and PS were supported by NSF grant SaTC~2228884. 
MS was supported by the DoD NDSEG Fellowship program and a MathWorks fellowship.

\appendix
\label{sec:appendix}

\section{Counterexamples}
\subsection{A UI-TS can be inconsistent}
\label{app:inconsistent_uits}

It is straightforward to find an inconsistent UI-TS defined by fixed selections $S_t(\bs{\eta}) := S_t$ and bets $\lambda_{kt}(\bs{\eta}) := \lambda \in (0, 1)$.
In this case, the set of I-TSMs $M_t(\bs{\eta})$ is log-concave over $\bs{\eta} \in \mathcal{C}$ (see \Cref{app_subsec:eta-oblivious-enumeration}), so the smallest I-TSM occurs in $\mathcal{V}$ (the vertices of $\mathcal{C}$). 
Thus the UI-TS is only consistent if with probability~1 all $\bs{\eta} \in \mathcal{V}$ are rejected.
Take $\bs{w} = [1/2,1/2]$ and $\eta_0 = 1/2$ so that
$\mathcal{V} = \{[0,1], [1,0]\}$, and let  
$x_{ik} = \mu^\star \in (1/2, 1)$ for all $i$ and $k$. 
Under $\bs{\eta} = [1,0]$, any sample from stratum~1 yields 
$
    a := Z_{1i}(\eta_1) = 1 - \lambda + \lambda \mu^\star < 1,
$ 
and any sample from stratum~2 yields 
$
    b := Z_{2i}(\eta_2) = 1 + \lambda \mu^\star > 1.
$
Therefore, after $T_1(t)$ draws from stratum~1 and $T_2(t)$ draws from stratum~2, we have
$
M_t([1,0]) = a^{T_1(t)} b^{T_2(t)}
$.
For $M_t([1,0])$ to reach $1/\alpha$ there must be some $t$ for which
$
T_2(t) \ge - T_1(t) (\nicefrac{\log a}{\log b}) - \nicefrac{\log \alpha}{\log b}.
$
Conversely, $M_t([0,1]) = b^{T_1(t)} a^{T_2(t)}$, so to reject $[0, 1]$ there must be some $t'$ for which
$
T_1(t') \geq - T_2(t') (\nicefrac{\log a}{\log b}) - \nicefrac{\log \alpha}{\log b}.$
Round-robin selection
$S_t := 1 + (t \mod 2)$ implies $|T_1(t) - T_2(t)| \leq 1$ for all $t$. 
The inequalities cannot both be satisfied for nontrivial $\alpha$: the UI-TS is inconsistent.

\subsection{Greedy Kelly I-TSMs may not be Kelly optimal for sampling without replacement}
\label{app:greedy_counterexample}
It is easy to show by counterexample that Greedy Kelly is not Kelly optimal for sampling without replacement from strata of different sizes. 
Indeed, let $K=2$, $N_1 = 10000$, and $N_2 = 3$.
Suppose the strata are point-masses $x_{ik} = \mu^*_k = 0.7$ for all $i$ and $k$. 
Take $\eta_1 = 0.49$ and $\eta_2 = 0.5$.
Returns are necessarily positive in each stratum so maximal bets ($\lambda_{ik} := 1 / \eta_{ik}$) are Kelly optimal.
In stratum~2, the conditional null mean after $i$ samples is  $\eta_{2i} = \frac{1.5 - 0.7(i-1)}{3-(i-1)}$, so the log growth at time 1 is $\log(0.7/0.5) \approx 0.34$, at time 2 is $\log(0.7/0.4) \approx 0.56$, and at time 3 is $\log(0.7/0.1) \approx 1.95$. 
On the other hand, the growth in stratum~1 is $\log(0.7/0.49) \approx 0.36$ initially, and will grow slowly with $t$.
Greedy Kelly would begin by drawing from stratum~1 and continue to do so until the stratum is exhausted, but the Kelly optimal strategy for horizons $t > 2$ would exhaust stratum~2 and \emph{then} sample stratum~1. 

\subsection{A marginally STO UI-TS may not be STO}
\label{app:jensen_gap}

Take $K = 2$, $\boldsymbol{w} = [1/2, 1/2]$, $\eta_0 = 1/2$ and suppose both strata are Bernoulli with $\mu^\star_1 = \mu^\star_2 =
p = 0.9$. 
Consider the two intersection nulls
$\boldsymbol\eta_A := [0.49,\, 0.51]$ and 
$\boldsymbol\eta_B := [0.51,\, 0.49]$.
The Kelly bet for Bernoullis is $\lambda^*(\eta) = (\mu^\star - \eta)/(\eta(1-\eta))$, which produces
maximal expected log-growth $g(\eta) := \mu^\star \log(1 + \lambda^*(\eta)(1-\eta)) + (1-\mu^\star)
\log(1 - \lambda^*(\eta)\eta)$. 
Here $g(0.49) \approx 0.384 > g(0.51) \approx
0.352$, so $M_t^*(\bs{\eta}_A)$ always draws from stratum~1, $M_t^*(\boldsymbol\eta_B)$
always draws from stratum~2, and their stopping times $\tau_A$ and $\tau_B$ are IID.
As a result, 
$$\mathbb{E}[\max(\tau_A, \tau_B)] = \mathbb{E}[\tfrac12 (\tau_A + \tau_B + |\tau_A - \tau_B|)]
  = \mathbb{E}[\tau_A] + \tfrac12 \mathbb{E}|\tau_A - \tau_B|.$$ 
The term $\mathbb{E}|\tau_A - \tau_B|$ measures the dispersion in the joint distribution of $(\tau_A, \tau_B)$, which can potentially be reduced by trading off for larger marginal means. 

Indeed, consider a competing UI-TS $M'_t$ that is identical at $\boldsymbol\eta_A$
but whose constituent at $\boldsymbol\eta_B$ \emph{also} draws only from
stratum~1.
Its marginal stopping time is strictly worse, since
$g(0.51) < g(0.49)$ implies $\mathbb{E}[\tau'_B] > \mathbb{E}[\tau_B]$. 
However, now both constituents are driven by the same data, so $\tau'_A$ and $\tau'_B$ are
comonotone and $\mathbb{E}|\tau'_A - \tau'_B|$ is relatively small.

Simulating $4 \times 10^5$ sample paths gives the
expected stopping times below:
\begin{center}
\begin{tabular}{lccc}
 & $\alpha = 0.05$ & $\alpha = 10^{-4}$ & $\alpha = 10^{-6}$ \\
 \hline
Greedy Kelly UI-TS $M^*_t$ & 10.42 & 29.18 & 42.45 \\
Competitor $M'_t$          & 9.58 & 27.14 & 39.99 \\
\hline 
Difference       & 0.85 & 2.04 & 2.46 \\
\end{tabular}
\end{center}

\section{Proofs}

\subsection{Proof of \Cref{lemma:sample_size_stopping_time}}
\label{app:proof_lemma_sample_size_stopping_time}
\noindent 
For Part 1, the first inequality follows from the fact that the sum of suprema is never less than the supremum of a sum; 
the second inequality follows from the fact that $\tau_k(M_t, \bs{\eta}) \leq \tau(M_t)$ for all $\bs{\eta} \in \mathcal{C}$, so $n_\tau(M_t) = \sum_{k=1}^K \sup_{\bs{\eta} \in \boundaryNull} \tau_k(M_t; \bs{\eta}) \leq \sum_{k=1}^K \tau(M_t) = K \times \tau(M_t)$. 
Part 2 is intuitive from the definitions. 
Technically it follows because, when every I-TSM has rejected using the same sequence of selections, the sampling depth in each stratum for any I-TSM $M_t(\bs{\eta})$ is upper bounded by the corresponding sampling depth of the last I-TSM to reject. 
This means $n_\tau(M_t) \leq \tau(M_t)$, so equality must hold. $\Box$ 

\subsection{Proof of \Cref{proposition:global_LCB}}
\label{app:proof_proposition_2}

Suppose that $\{L_{k t}\}_{k=1}^K$ are simultaneously and sequentially valid LCBs for 
$\{\mu_k\}_{k=1}^K$,
and that $\popN \in \nullDist$. 
Then
\begin{align*}
     \Pr_{\popN} (\exists~ t \in \mathbb{N} : L_t > \mu) &= \Pr_{\popN} \left (\exists~ t \in \mathbb{N} : \sum_{k=1}^K w_k L_{k T_k(t)} > \sum_{k=1}^K w_k \mu_k\right )\\ 
     &\leq \Pr_{\popN} \left (\exists~ t \in \mathbb{N} : \bigcup_{k=1}^K L_{k T_k(t)} > \mu_k \right )  \leq \alpha.
\end{align*}
The second step follows via $\left \{ \sum_{k=1}^K w_k L_{k T_k(t)} > \sum_{k=1}^K w_k \mu_k \right \} \implies \left\{ \bigcup_{k=1}^K L_{k T_k(t)} > \mu_k \right \}$. 
The final step follows from simultaneous validity of the LCBs. $\Box$

\subsection{Proof of \Cref{lem:kelly-optimal_I-TSM}}
\label{app:kelly-optimal-TSM}

The expected log-growth of I-TSM $M_t(\bs{\eta})$ can be written 
$\EE \left [ \Delta M_t(\bs{\eta}) \right ] = \sum_{k=1}^K p_{kt}(\eta_k) \EE[\log Z_{kt}(\eta_k)],$
where the expectation is with respect to the sampling and the stratum selections.
For any fixed selection strategy $(\bs{p}_t(\bs{\eta}))_{t \in \mathbb{N}}$, the overall expected log-growth is an increasing function of the expected-log growth within strata, and is maximized when each $\EE[\log Z_{kt}(\eta_k)]$ is maximized. 
This is ensured by using the Kelly optimal bet $\lambda_k^* = \argmax_{\lambda \in [0,1/\eta_k]} \EE[\log Z_{kt}(\eta_k)]$ within each stratum. 
Now, choose the bets $\bs{\lambda}^*$ so that each term $\EE[\log Z_{kt}(\eta_k)]$ is maximized. 
The selection probabilities
$
p_{k}^*(\bs{\eta}) := 1\{k = \argmax_j \EE[\log Z_{jt}(\eta_j)] \}
$ 
maximize $\EE[\Delta M_t(\bs{\eta})]$ because they always draw from a stratum with the largest expected log growth. $\Box$

\section{Dynamic program for sampling without replacement}
\label{app:stochastic_dp}

Fix an intersection null $\bs{\eta}$, a simple alternative $\popN \in \altDist$, and a time horizon $t$.
Under sampling without replacement, define the \emph{state at time $i$} by
$\omega_i := (X_1^{T_1(i)}, \ldots, X_K^{T_K(i)})$. 
The next draw from stratum $k$ is uniform on $\mathcal{R}_k(\omega_i) := \pop_k \setminus X_k^{T_k(i)}$, the $N_k - T_k(i)$ remaining values in $\pop_k$. 
The optimal conditional expected log-growth in stratum $k$ at time $i$ is
$$
g_k^*(\omega_i) := \max_{\lambda \in [0, 1/\eta_{ki}]} \frac{1}{|\mathcal{R}_k(\omega_i)|} \sum_{x \in \mathcal{R}_k(\omega_i)} \log\big(1 + \lambda(x - \eta_{ki})\big),
$$
attained by the greedy Kelly bet $\lambda_{ki}^*(\bs{\eta})$ of \Cref{lem:kelly-optimal_I-TSM}.
To maximize $\EE_{\popN}[\log M_t(\bs{\eta})]$, define the \emph{value-to-go} $V_j(\omega)$ from a generic state $\omega$ when there are $j$ draws remaining:
$$
V_0(\omega) := 0, \qquad V_j(\omega) := \max_{k \,:\, T_k(i) < N_k} \Big\{
\underbrace{g_k^*(\omega)}_{\substack{\text{growth collected} \\ \text{now from stratum } k}}
+ \underbrace{\EE\big[V_{j-1}(\Phi_k(\omega))\big]}_{\substack{\text{value-to-go from the next} \\ \text{state, averaged over draws}}}
\Big\},
$$
where $\Phi_k(\omega)$ is the state after one additional draw from stratum $k$.
At time 0, the state $\omega_0$ is empty and the value to go is $V_t(\omega_0) = \sup \EE_{\popN}[\log M_t(\bs{\eta})]$, with the supremum taken over predictable bets and selections.
The \emph{DP policy} $\pi^*$ pairs the greedy Kelly bets $\lambda_{ki}^*(\bs{\eta})$ with an \emph{optimal selection rule} $S_i^*$, which selects a stratum attaining the maximum in the recursion; write $\omega_i^*$ for the states it generates.
$\pi^*$ attains $V_t(\omega_0)$, so it is Kelly optimal.
To see this, let $M_t^{\pi}$ be generated by any strategy $\pi$, and consider the augmented process $U_i^{\pi} := \log M_i^{\pi} + V_{t-i}(\omega_i)$, which adds the value-to-go to the current log-wealth. 
Note $\mathbb{E}[U_0^\pi] = U_0^\pi = V_t(\omega_0)$ and $U_t^{\pi} = \log M_t^{\pi}$.
Conditioning on $\mathcal{F}_{i-1}(\bs{\eta})$ and using $\EE[\log \tilde{Z}_i \mid \mathcal{F}_{i-1}(\bs{\eta})] \leq g_{S_i}^*(\omega_{i-1})$ with the maximum above defining $V$ gives $\EE[U_i^{\pi} \mid \mathcal{F}_{i-1}(\bs{\eta})] \leq U_{i-1}^\pi$, so the augmented process is a supermartingale starting at the optimal value: $\EE_{\popN}[\log M_t^{\pi}] = \EE[U_t^\pi] \leq U_0^\pi = V_t(\omega_0)$. 
Under $\pi^*$ both inequalities are equalities at every step: the Kelly bet $\lambda_{ki}^*(\bs{\eta})$ attains $g_{S_i}^*(\omega_{i-1})$, and the selection $S_i^*$ attains the maximum defining $V$.
Hence $(U_i^{\pi^*})$ is a martingale along $(\omega_i^*)$: its expectation is constant and $\EE_{\popN}[\log M_t^{\pi^*}] = \EE[U_t^{\pi^*}] = U_0^{\pi^*} = V_t(\omega_0)$.
Unlike the greedy rule, the Kelly optimal selection rule draws from the stratum maximizing immediate conditional log-growth \emph{plus} the expected value-to-go of the resulting state, forgoing immediate gain to drain strata whose depletion unlocks higher growth or certain rejection later.

\section{Computation}
\label{app:computation}

\subsection{Brute-force search when the population and support are small}

Let $\mathcal{U}_k$ denote the possible values of elements of stratum $k$.
In general, $\mathcal{U}_k \subseteq [0,1]$ and $|\mathcal{U}_k|$ may be uncountably infinite.
However, when $|\mathcal{U}_k|$ is finite, the number of possible stratum means $|\Omega_k|$ is also finite. 
If this holds for all $k$, then $\mathcal{B}$ may be small enough to enumerate, enabling exhaustive minimization. 
This was the approach taken in \citet{SpertusStark2022}, but $|\Omega_k|$ is usually too large. 
For instance, when the strata are of equal size ($\bs{w} = K^{-1} \bs{1}$) and binary ($\mathcal{U}_k = \{0,1\}$), the number of possible vectors of stratum means is:
$
(N_k + 1)^K = (N/K + 1)^K.
$

\subsection{Log-concavity}
\label{app_subsec:eta-oblivious-enumeration}

If 
$\bs{\lambda}_t(\bs{\eta}) := \bs{\lambda}_t$ 
and 
$\bs{p}_t(\bs{\eta}) := \bs{p}_t$ 
are $\bs{\eta}$-oblivious,
$M_t(\bs{\eta})$ is log-concave in $\bs{\eta}$. 
The second partial derivative of $\log M_t(\bs{\eta})$ with respect to $\eta_k$ is:
\begin{equation*}
    \frac{\partial^2}{\partial \eta_k^2} \log M_t(\bs{\eta}) = - \sum_{i=1}^{\Tkt{k}{t}} \left [ \frac{\lambda_{ki}}{1 + \lambda_{ki} (X_{ki} - \eta_k)}\right ]^2 \geq 0.
\end{equation*}
The mixed partials vanish. 
Thus, the Hessian of $\log M_t(\bs{\eta})$ is negative semi-definite and $\log M_t(\bs{\eta})$ is concave in $\bs{\eta}$.
Since $M_t(\bs{\eta})$ is log-concave, it is quasi-concave: its value on any line segment within $\boundaryNull$ is not smaller than the smaller of its values at the ends of the segment.
Since only vertices $\mathcal{V}$ are not strict convex combinations of other points in $\boundaryNull$, the minimum over $\boundaryNull$ is attained in $\mathcal{V}$.

\begin{algorithm}
\caption{Banded UI-TS under stratified sampling with replacement when $K=2$}
\label{alg:uits}
\begin{algorithmic}[1]  
\Require stratified sample $((X_{kt})_{t \in \mathbb{N}})_{k=1}^K$, selection rules $(\bs{p}_t(\bs{\eta}))_{t \in \mathbb{N}, \bs{\eta} \in \mathcal{C}}$, betting strategies $(\bs{\lambda}_t(\bs{\eta}))_{t \in \mathbb{N}, \bs{\eta} \in \mathcal{C}}$, level $\alpha$, global null $\eta_0$, stratum weights $\bs{w}$, and grid size $G$
\Ensure Stopping time and sample size of betting UI-TS $M_t$ for hypothesis $H_0: \mu^* \leq \eta_0$
\State Partition $\{\bs{\eta}: \bs{w} \cdot \bs{\eta} = \eta_0,  \bs{\eta}  \in [0,1]\}$ into $G$ equal-length line segments with vertices $(\bs{v}_{1g}, \bs{v}_{2g})$, where $v_{1gj}$ represents the mean in the $j$th stratum at the vertex of the $g$th of the band. 
\State For each band $g \in [G]$, compute its centroid $\bar{\bs{\mu}}_g = (v_{2g} - v_{1g})/2$ for $g \in [G]$ and an upper bound on the stratum-wise mean $\tilde{\bs{\mu}}_g = [\max \{v_{1g1}, v_{2g1}\} , \max \{v_{1g2}, v_{1g2}\} ]$. 
\State Initialize: time $t \gets 1$, log-UI-TS $M_0 \gets 0$, running sample sizes $\bs{T}_{0g} \gets [0,0]$, selection $S_{tg} \sim \mbox{Categorical}([0.5,0.5])$.

\While{$U_t < 1/\alpha$}
    \For{$g \in [G]$} 
        \State Compute selection $S_{tg} \sim \mbox{Categorical}[\bs{p}_t(\bar{\mu}_g)]$
        \State Update sample size $\bs{T}_{tg} \gets \bs{T}_{tg} + \bs{e}^{(S_{tg})}$
        \State Compute I-TSMs at vertices: $M_{t}(v_{jg}) = \prod_{k=1}^K \prod_{i=1}^t [1 + \lambda_{tk}(\tilde{\mu}_g) (X_{T_{tgk}k} - v_{jg})$ 
        \State Compute smallest I-TSM in band: $M_{t g} \gets \min\{M_t(v_{1g}), M_t(v_{2g})\}$
        \EndFor
        
    \State $U_{t} \gets \min_{g \in [G]} M_{tg}$
    \State $t \gets t + 1$
\EndWhile
\State $\tau \gets t$
\State $n_\tau \gets \max_{g\in[G]} T_{1 \tau g} + \max_{g \in [G]} T_{2 \tau g} $
\State \Return $(\tau, n_\tau)$
\end{algorithmic}
\end{algorithm}

\section{Comparison to PEAK}
\label{app:peak-comparison}

We compare our UI-TS (AGRAPA bets, $G=50$ bands) and the LCB method (Section~3.1, Šidák-corrected, AGRAPA
bets) to PEAK, which uses an average-combined UI-TS with a special betting rule \citep{cho_peeking_2024}. 
PEAK results in a convex objective, which we minimize numerically over a grid of 250 points. 
This technique is not strictly valid as the true minimum may not be attained, but it upper bounds PEAK's efficiency. 
All three methods use round-robin sampling and $\alpha = 0.05$.

We consider two scenarios with $K=2$ equally-sized strata and $\eta_0 = 1/2$: (i) point
masses with $\bs{N} = [200,200]$, global reported assorter mean $\bar A^c$ ranging
from 0.51 to 0.8, and a gap between stratum-wise assorter means of 0 or 0.2;
and (ii) Bernoulli populations as in
Section~6.2, with $\bs{N} =[600,600]$ and mean $\mu^\star = 0.6$ in both strata.
The Bernoulli simulations were replicated 100 times. 
Table~\ref{tab:peak-comparison} reports mean and median sample size for each method across simulations and settings, capped at $N$ samples.
UI-TS is uniformly most efficient. 
PEAK is comparable to the LCB method for Bernoullis, but substantially less efficient than both product-combined UI-TS and LCB method for the point-mass populations.
We expect the gap will become more pronounced for larger $K$, as product combining makes increasing gains over averaging.

\begin{table}[ht]
\centering
\begin{minipage}{0.48\textwidth}
\centering
\begin{tabular}{lcc}
\hline
Method & Mean & Median \\
\hline
UI-TS & 62.7 & 15.5 \\
LCB   & 79.2 & 32.5 \\
PEAK  & 157.3 & 80.0 \\
\hline
\end{tabular}
\caption{Point masses (comparison RLA): $N=[200,200]$, $\bar A^c \in [0.51,0.8]$, stratum gap $\in \{0, 0.2\}$.}
\label{tab:peak-pointmass}
\end{minipage}
\hfill
\begin{minipage}{0.48\textwidth}
\centering
\begin{tabular}{lcc}
\hline
Method & Mean & Median \\
\hline
UI-TS & 393.5 & 375.5 \\
LCB   & 616.5 & 585.0 \\
PEAK  & 546.5 & 484.0 \\
\hline
\end{tabular}
\caption{Bernoulli: $N=[600,600]$, $\mu_k^\star = 0.6$ in both strata, 100 replicates.}
\label{tab:peak-bernoulli}
\end{minipage}
\caption{Mean and median global sample size ($n_\tau$) under our product-combined union-of-intersections test sequence (UI-TS), combined lower confidence bounds (LCB), and PEAK \citep{cho_peeking_2024}.}
\label{tab:peak-comparison}
\end{table}

\bibliography{pbsBib}

\end{document}